# Non-Markovian noise sources for quantum error mitigation

Doyeol Ahn[1,2,*] and Byeongyong Park[1,3]

[1]Department of Electrical and Computer Engineering,
University of Seoul, 163 Seoulsiripdae-ro, Tongdaimoon-gu, Seoul 02504, Republic of Korea

[2]First Quantum, Inc, #210, 180 Bangbae-ro, Seocho-gu, Seoul 06586, Republic of Korea

*Corresponding author: dahn@uos.ac.kr

**ABSTRACT**

Reducing the impact of errors and decoherence in near-term quantum computers, such as noisy intermediate-scale quantum (NISQ) devices, is critical for their practical implementation. These factors significantly limit the applicability of quantum algorithms, necessitating a comprehensive understanding of their physical origins to establish effective error mitigation strategies. In this study, we present a non-Markovian model of quantum state evolution and a quantum error mitigation cost function tailored for NISQ devices interacting with an environment represented by a set of simple harmonic oscillators as a noise source.

Employing the projection operator formalism and both advanced and retarded propagators in time, we derive the reduced-density operator for the output quantum states in a time-convolutionless form by solving the quantum Liouville equation. We examine the output quantum state fluctuations for both identity and controlled-NOT (CNOT) gate operations in two-qubit operations using a range of input states. Subsequently, these results are compared with experimental data from ion-trap and superconducting quantum computing systems to estimate the crucial parameters of the cost functions for quantum error mitigation.

Our findings reveal that the cost function for quantum error mitigation increases as the coupling strength between the quantum system and its environment intensifies. This study underscores the significance of non-Markovian models in understanding quantum state evolution and highlights the practical implications of the quantum error mitigation cost function when assessing experimental results from NISQ devices.

**Introduction**

Investigating decoherence and gate errors as sources of noise is vital for understanding the underlying physical processes in quantum computing systems. These noise sources impose time and size constraints for practical applications [1]. Despite these limitations, shallow quantum circuits hold potential significance for solving practical problems through hybrid quantum-classical algorithms [2,3]. These hybrid systems can produce significant results that surpass classical computers, even with finite error rates, due to their error-resilience [4,5]. To maximize error suppression or mitigation while minimizing the number of qubits, it is essential to consider the high cost of full-scale fault-tolerant quantum computing [6-8]. Noise arises from interactions between the qubit system and its environment, making it crucial to comprehend the physical basis of these errors for effective error mitigation strategy implementation.

The current Quantum Error Mitigation (QEM) scheme necessitates information on noise parameters, system size, and circuit depth [6]. QEM methods often involve gate models to describe quantum circuits and time-dependent Hamiltonian dynamics, incorporating the Lindblad operator as a noise source [6]. Descriptions of noise sources related to quantum state fluctuations can be derived from the kinetic perspective of non-equilibrium quantum statistical mechanics and are rooted in the quantum Liouville equation [9]. In many cases, conventional Langevin-type descriptions of these fluctuations result in memoryless noise sources, known as Markovian events [9].

The primary focus in studying noise sources affecting Noisy Intermediate-Scale Quantum (NISQ) computers has been on Markovian noise. However, it is crucial to examine the relationship between quantum error mitigation and non-Markovian noise sources, as these noise effects are practically unavoidable in most solid-state devices used for quantum computing. Markovian noise is characterized by memoryless behavior, while non-Markovian noise sources exhibit memory effects, where noise at any given moment depends on the system's past history [10-12]. Non-Markovian noise is more complicated to model and presents more demanding error mitigation challenges.

In this study, we present a non-Markovian approach to analyze quantum state fluctuations in near-term quantum computing devices, aiming to determine the Quantum Error Mitigation (QEM) cost function. This approach is relevant as non-Markovian noise effects are common in these devices, and understanding their impact is essential for developing effective error mitigation strategies. Employing the non-Markovian approach and modeling the environment with simple harmonic oscillators allows us to investigate the relationship between noise sources and the quantum system's performance. This understanding is crucial for determining the QEM cost function, which quantifies the effectiveness of error reduction techniques applied to a quantum computing system.

The insights gained from this non-Markovian approach can inform tailored quantum error mitigation strategies for near-term quantum computing devices. Addressing the unique challenges posed by non-Markovian noise sources enables researchers to design more robust and efficient error mitigation techniques, leading to improved performance and reliability of these devices in the presence of various environmental disturbances.

Dissipation in quantum dynamics of a two-state or qubit system is widely encountered in physics and chemistry, and its description remains a significant theoretical challenge, especially when considering non-Markovian processes [10-13]. These non-Markovian processes related to dissipation are closely linked to the low-frequency noise spectrum [14-16]. It has been suggested that low-frequency noise plays a crucial role in decoherence processes in ion trap or superconducting systems for near-term quantum computers [3]. Understanding non-Markovian quantum dynamics of a two-state system is also important for evaluating the performance of an adiabatic quantum computer in the presence of noise [17]. This is because, for many difficult problems, the computation bottleneck involves passing through a point where the gap between the ground state and the first excited state is small, and the system is strongly coupled to the environment. One of the authors previously developed a time-convolutionless reduced-density operator theory of noisy quantum-state evolution [18, 19]. This time-convolutionless formulation was shown to incorporate both non-Markovian relaxation and memory effects [20]. The time-convolutionless equations of motion were initially suggested by Tokuyama and Mori in the Heisenberg picture [21], and later developed in the Schrödinger picture using the projection operator technique [22-23]. In this work, we extend the reduced-density operator theory to derive an expression for the quantum Liouville equation for a two-state system, making it more suitable for perturbational analysis to obtain the master equation in the presence of low-frequency noise and strong coupling to the environment.

This extended approach enables a comprehensive exploration of non-Markovian noise impacts on quantum computing device performance. By analyzing various noise sources' influence on system dynamics, researchers can pinpoint major error contributors and devise mitigation strategies. Moreover, understanding non-Markovian noise characteristics facilitates tailored error correction techniques addressing these sources' unique challenges. Studying non-Markovian noise sources and their quantum error mitigation relationship is crucial for quantum computing technology advancement. Utilizing a non-Markovian approach to assess quantum state fluctuations and addressing the distinct challenges of these noise sources allows for more effective error mitigation strategies, enhancing near-term quantum computing device performance and reliability. This work ultimately contributes to harnessing quantum computing power for numerous practical applications, including cryptography and materials science.

**Theoretical formulation**

The total Hamiltonian for an open two-state system is given by [12, 18,19, 25]

$$\widehat{H}_T = \widehat{H}_S(t) + \widehat{H}_B + \widehat{H}_{int} \tag{1}$$

where $\widehat{H}_S(t)$ is the system Hamiltonian for a two-state system, $\widehat{H}_B$ the Hamiltonian acting on the reservoir or an environment, $\widehat{H}_{int}$ is the interaction between the system and the environment (as depicted in figure 1).

The equation of motion for the total density operator $\rho_T(t)$ of the total system is given by a quantum Liouville equation [18]

$$\frac{d}{dt}\rho_T(t) = -i[\widehat{H}_T(t), \rho_T(t)] = -i\widehat{L}_T(t)\rho_T(t), \tag{2}$$

where

$$\widehat{L}_T(t) = \widehat{L}_S(t) + \widehat{L}_B(t) + \widehat{L}_{int}(t) \tag{3}$$

is the Liouville super operator in one-to-one correspondence with the Hamiltonian. Here, we use the unit in which $\hbar = 1$. In order to derive an equation and to solve for a system alone, it is convenient to use the projection operators which decompose the total system by eliminating the degrees of freedom for the reservoir. We define thine-independent projection operator $\underline{P}$ and $\underline{Q}$ given by [18]

$$\underline{P}X = \rho_B tr_B X, \underline{Q} = 1 - \underline{P} \tag{4}$$

for any dynamical variable $X$. Here $tr_B$ denotes a partial trace over the quantum reservoir and $\rho_B$ is the density matrix of the reservoir. The projection operators satisfy the following operator identities:

$$\underline{P}^2 = \underline{P}, \ \underline{Q}^2 = \underline{Q} \ \underline{P}\,\underline{Q} = \underline{Q}\underline{P} = 0, \tag{5a}$$

$$\underline{P}\widehat{L}_S(t) = \widehat{L}_S(t)\underline{P}, \ \underline{P}\widehat{L}_B = \widehat{L}_B\underline{P} = 0, \ \underline{P}\widehat{L}_{int}\underline{P} = 0, \tag{5b}$$

and $\quad \underline{Q}\widehat{L}_S(t) = \widehat{L}_S(t)\underline{Q}, \ \underline{Q}\widehat{L}_B = \widehat{L}_B\underline{Q} = \widehat{L}_B. \tag{5c}$

The information of the system is then contained in the reduced density operator

$$\rho(t) = tr_B \rho_T(t) = tr_B \underline{P}\rho_T(t). \tag{6}$$

If we multiply equation (2) by $\underline{P}$ and $\underline{Q}$ on the left side, we obtain coupled equations for $\underline{P}\rho_T(t)$ and $\underline{Q}\rho_T(t)$ as follows:

$$\frac{d}{dt}\underline{P}\rho_T(t) = -i\underline{P}\widehat{L}_T(t)\underline{P}\rho_T(t) - -i\underline{P}\widehat{L}_T(t)\underline{Q}\rho_T(t), \tag{7a}$$

and

$$\frac{d}{dt}\underline{Q}\rho_T(t) = -i\underline{Q}\widehat{L}_T(t)\underline{Q}\rho_T(t) - -i\underline{Q}\widehat{L}_T(t)\underline{P}\rho_T(t), \tag{7b}$$

where we have modified the Eq. (2) as

$$\frac{d}{dt}\rho_T(t) = -i\widehat{L}_T(t)\rho_T(t) = --i\widehat{L}_T(t)(\underline{P}+\underline{Q})\rho_T(t).$$

We assume that the system was turned on at $t = 0$ and the input state prepared at $t = 0$ was isolated with the reservoir such that $\underline{Q}\rho_T(0) = 0$.

The formal solution of Eq. (2) is given by (Supplementary Information)

$$\rho_B \rho(t) = \rho_B \widehat{U}_S(t,0)\rho(0) - i\rho_B \int_0^t ds \widehat{U}_S(t,s) tr_B\left[\widehat{L}_T(s)\{\widehat{\theta}(s)-1\}\rho_B\right] tr_B\left[\underline{G}(t,s)\widehat{\theta}(t)\rho_B\right]\rho(t), \quad (9)$$

or

$$\left(1 + i\int_0^t ds \widehat{U}_S(t,s) tr_B\left[\widehat{L}_T(s)\{\widehat{\theta}(s)-1\}\rho_B\right] tr_B\left[\underline{G}(t,s)\widehat{\theta}(t)\rho_B\right]\right)\rho(t) \quad (10)$$
$$= \widehat{U}_S(t,s)\rho(t).$$

Here $\widehat{U}_S(t,s)$ is the propagator of the system defined by

$$\widehat{U}(t,s)\underline{P} = \underline{T}exp\left[-i\int_s^t d\tau \underline{P}\widehat{L}_T(\tau)\underline{P}\right]\underline{P}$$
$$= \underline{T}exp\left[-i\int_s^t d\tau \widehat{L}_S(\tau)\underline{P}\right]\underline{P} \quad (11)$$
$$= \underline{T}exp\left[-i\int_s^t d\tau \widehat{L}_S(\tau)\right]\underline{P} = \widehat{U}_S(t,s)\underline{P}.$$

If we define $\widehat{W}(t)$ by

$$\widehat{W}(t) = 1 + i\int_0^t ds \widehat{U}_S(t,s) tr_B\left[\widehat{L}_T(s)\{\widehat{\theta}(s)-1\}\rho_B\right] tr_B\left[\underline{G}(t,s)\widehat{\theta}(t)\rho_B\right], \quad (12)$$

Then, the evolution operator for the reduced density operator $\rho(t)$ is given by

$$\rho(t) = \widehat{W}^{-1}(t)\widehat{U}_S(t,0)\rho(0) = \widehat{V}(t)\rho(0), \quad (13)$$

where the super-operator $\widehat{V}(t)$ for the evolution of the reduced density operator is defined by $\widehat{V}(t) = \widehat{W}^{-1}(t)\widehat{U}_S(t,0)$. Within the Born approximation, we have (Supplementary Information)

$$\widehat{V}^{(2)}(t) = \left\{1 - \int_0^t ds \int_0^s d\tau \widehat{U}_S(t,s) tr_B\left[\widehat{L}_{int}\widehat{U}_o(s,\tau)\widehat{L}_{int}\widehat{U}_o^{-1}(s,\tau)\rho_B\right]\widehat{U}_o^{-1}(t,s)\right\}\widehat{U}_S(t,0), \quad (14)$$

where

$$\widehat{U}_o(t) = \underline{T}exp\left[-i\int_0^t d\tau\left(\widehat{L}_S(\tau) + \widehat{L}_B\right)\right]$$
$$= exp\left(-it\widehat{L}_B\right)\underline{T}exp\left[-i\int_0^t d\tau\left(\widehat{L}_S(\tau)\right)\right] \quad (15)$$
$$= \widehat{U}_B(t)\widehat{U}_S(t).$$

After some mathematical manipulations, $\widehat{V}^{(2)}(t)\rho(0)$ becomes

$$\rho(t)$$
$$= \widehat{V}^{(2)}(t)\rho(0)$$
$$= \widehat{U}_S(t)\rho(0) - \int_0^t ds \int_0^s d\tau\, tr_B\left[\widehat{H}_{int}\widehat{H}_{int}(\tau-s)\rho_B\rho(-s)\right] + \int_0^t ds \int_0^s d\tau\, tr_B\left[\widehat{H}_{int}(\tau-s)\rho_B\rho(-s)\rho_B\widehat{H}_{int}\right]$$
$$+ \int_0^t ds \int_0^s d\tau\, tr_B\left[\widehat{H}_{int}\rho_B\rho(-s)\widehat{H}_{int}(\tau-s)\right] - \int_0^t ds \int_0^s d\tau\, tr_B\left[\rho_B\rho(-s)\widehat{H}_{int}(\tau-s)\widehat{H}_{int}\right].$$
$$(16)$$

Here $\widehat{H}_{int}(\tau-s)$ and $\rho(-s)$ are Heisenberg operators defined by

$$\widehat{H}_{int}(\tau-s) = \exp\left(i\widehat{H}_{int}(\tau-s)\right)\widehat{H}_{int}\exp\left(-i\widehat{H}_{int}(\tau-s)\right),$$

$$\rho(-s) = \exp\left(-i\widehat{H}_S s\right)\rho(0)\exp\left(i\widehat{H}_S s\right),$$

respectively.

**Results**

In this work, we focus on the two-qubit gate operations and model the interaction of the quantum system with the environment during the gate operation by a Caldeira-Leggett model [11-13, 25] where a set of harmonic oscillators are coupled linearly with the system spin by

$$\widehat{H}_{int} = \lambda \sum_{i=1,2,\, j=1,2,3} \vec{S}_i \cdot \vec{b}_i, \quad S_i^j = \frac{\hbar}{2}\sigma^j \tag{17}$$

where $\sigma_i$ is the Pauli matrices $\sigma_1 = X$, $\sigma_2 = Y$, $\sigma_3 = Z$ and $b_i^j$ is the fluctuating quantum field associated the ith qubit, whose motion is governed by the harmonic-oscillator Hamiltonian. In the evaluation of Eq. (16), we obtain the following relations:

$$tr_B\{b_k^l(t) b_i^j \rho_B\} = \Gamma(t) + i\Delta(t),$$
$$tr_B\{b_i^j b_k^l(t) \rho_B\} = \Gamma(t) - i\Delta(t), \tag{18}$$

where

$$\Gamma(t) + i\Delta(t) = \frac{\lambda^2}{\pi}\int_0^\infty J(\omega)\left\{\exp(-i\omega t) + \coth\left(\frac{\omega}{2k_B T}\right)\cos\omega t\right\}. \tag{19}$$

Here $J(\omega)$ is the ohmic damping given by $J(\omega) = \theta(\omega_c - \omega)\eta\omega$, $\Gamma$ is the decoherence rate of the qubit system.

We evaluate the reduced-density-operator in the multiplet basis representation [18]

$$\rho(t) = \sum_{a,b} \rho_{ab}(t) e_{ab}, \quad e_{ab} = |a\rangle\langle b|, \quad a,b = 1,2,3,4, \tag{20}$$

where $e_{ab}$ is the multiplet states. The inner product between the multiplet basis is defined by

$$(e_{ab}, e_{cd}) = tr\left[e_{ab}^\dagger e_{cd}\right] = \delta_{ac}\delta_{bd}. \tag{21}$$

Then, from Eqs. (16)-(21), we obtain the matrix component of the reduced-density-operator as (Supplementary Information)

$$\rho_{ab}(t) = V_{ab|cd}(t)\rho(0),$$

$$V_{ab|cd}(t) = \exp\left[-it(E_a - E_b)\right]\left\{\delta_{ac}\delta_{bd} - \left[\delta_{bd}\sum_{a'} M_{aa'a'c} - M_{acdb}\right]k(t) - \left[\delta_{ac}\sum_{a'} M_{aa'a'b} - M_{acdb}\right]k^*(t)\right\} \tag{22}$$

where

$$M_{abcd} = \sum_{i,j} \langle a|S_i^j|b\rangle\langle c|S_i^j|d\rangle$$
$$= \frac{1}{4}\sum_{i=1,2}\left\{\langle a|X_i|b\rangle\langle c|X_i|d\rangle + \langle a|Y_i|b\rangle\langle c|Y_i|d\rangle + \langle a|Z_i|b\rangle\langle c|Z_i|d\rangle\right\}, \tag{23}$$

and
$$k(t) = \frac{2}{\pi}\Gamma_o\left\{\frac{\pi}{2}\omega_c t + \int_0^{\frac{t}{\tau_s}} Si(\omega_c \tau_s t) dt\right\} \quad (24)$$
$$+ i\Delta_o\left\{\frac{t}{\tau_s} - \frac{1}{\omega_c \tau_s}\left(\frac{\pi}{2}\omega_c \tau_s + Si(\omega_c t)\right)\right\}.$$

Here $\omega_c$ is the high frequency cutoff, $\tau$ is the switching time, $\Gamma_0 = \lambda^2 \eta k_B T \tau_s$ and $\Delta_o = \frac{\lambda^2 \eta \omega_c \tau_s}{\pi}$.

We now study the non-Markovian errors associated with two-qubit gate operations. In Eq. (24),

$Si(x) = \int_0^x dt \frac{\sin t}{t} - \frac{\pi}{2}$, is a sine integral [27].

In Eq. (22), $V_{ab|cd}(t)$ is the quantum evolution operator for the reduced density operator containing the effects of non-Markovian noise sources. The purpose of QEM is to restore the ideal quantum evolution without noisy processes. If we denote the ideal quantum volution and QEM based recovery operation as $\varepsilon_I(t)$ and $R_{QEM}(t)$, respectively, such that

$$\varepsilon_{I\,ab|cd}(t) = \sum_{e,f} R_{QEM\,ab|ef}(t) V_{ef|cd}(t), \quad (25)$$

and
$$R_{QEM}(t) = \sum_i \mu_i O_i = c(t) \sum_i sgn(\mu_i) p_i O_i. \quad (26)$$

Here, $c(t) = \sum_i |\mu_i|$ is the QEM cost function, $p_i = \frac{|\mu_i|}{c(t)}$, and $\{O_i\}$ is the set of physical operations applied for QEM.

We first consider the identity operation depicted in figure 2.

For identity operation, the multiple basis is given by

$$|1>^m = |00>,$$
$$|2>^m = \frac{1}{\sqrt{2}}(|01> + |10>),$$
$$|3>^m = |11>, \quad (27)$$
$$|4>^m = \frac{1}{\sqrt{2}}(|01> - |10>)$$

Assuming that the initial state is given by $\rho_{cd}(0) = |00><00| = \rho_{11}^m(0)$, we obtain

$$\rho_{11}^{m}(t) = V_{11|11}(t)\rho_{11}^{m}(0)$$

$$= \left\{1 - 2\left[\sum_{a'} M_{1a'a'1} - M_{1111}\right] Re\, k(t)\right\} \rho_{11}^{m}(0)$$

$$= (1 - 2\, Re\, k(t))\rho_{11}^{m}(0),$$

$$\rho_{22}^{m}(t) = V_{22|11}(t)\rho_{11}^{m}(0)$$

$$= 2M_{2112} Re\, k(t) \rho_{11}^{m}(0)$$

$$= Re\, k(t) \rho_{11}^{m}(0),$$

$$\rho_{33}^{m}(t) = V_{33|11}(t)\rho_{11}^{m}(0)$$

$$= 2M_{3113} Re\, k(t) \rho_{11}^{m}(0)$$

$$= 0,$$

$$\rho_{44}^{m}(t) = V_{44|11}(t)\rho_{11}^{m}(0)$$

$$= 2M_{4114} Re\, k(t) \rho_{11}^{m}(0)$$

$$= Re\, k(t) \rho_{11}^{m}(0). \tag{28}$$

In NISQ machines, the input and out states are represented by the computational basis:

$$e_{\alpha\beta}^{c} = |\alpha\rangle\langle\beta|, \quad \alpha,\beta = 1,2,3,4, \quad \{|00\rangle, |01\rangle, |10\rangle, |11\rangle\}. \tag{29}$$

Then the reduce-density-operator in the computational basis is given by

$$\rho_{\alpha\beta}^{c}(t) = \sum_{a',b'} C_{\alpha\beta|a'b'} \rho_{a'b'}^{m}(t),$$

$$C_{\alpha\beta|ab} = tr\left(e_{\alpha\beta}^{c\dagger}, e_{ab}^{m}\right). \tag{30}$$

By substituting, Eqs. (38). (39) into Eq. (36), we obtain '

$$\rho_{11}^{c}(t) = (1 - 2\, Re\, k(t))\rho_{11}^{m}(0),$$

$$\rho_{22}^{c}(t) = Re\, k(t) \rho_{11}^{m}(0),$$

$$\rho_{33}^{c}(t) = Re\, k(t) \rho_{11}^{m}(0)$$

$$\rho_{44}^{c}(t) = 0. \tag{31}$$

Here $1 - 2Re\, k(t)$, $Re\, k(t)$, $Re\, k(t)$, $0$ are the probabilities of finding the output states of the system in $|00\rangle, |01\rangle, |10\rangle, |11\rangle$ states at time $t$, respectively.

In figure 3, we show the plot of $Re\, k(t)$ vs $\frac{t}{\tau_s}$, where $\tau_s$ is the switching time for the parameters $\Gamma_0 \omega_c \tau_s$ in the range $7.0 \times 10^{-4} \leq \Gamma_0 \omega_c \tau_s \leq 7.0 \times 10^{-3}$.

To estimate the quantum state fluctuation in practical NISQ devices, two platforms were employed: ibm_guadalupe through IBM Quantum and ionQ through Amazon Braket. The probabilities of each state after 1,000 iterations are presented in Table 1.

Table 1: Quantum state fluctuations under the identity operation

| Initial state: \|00> | \|00> | \|01> | \|10> | \|11> |
|---|---|---|---|---|
| Ibm_guadalupe | 0.987 | 0.006 | 0.007 | 0 |
| IonQ | 0.998 | 0.001 | 0.001 | 0 |
| Theory: probability for each iteration at time t | $1 - 2 Re\, k(t)$ | $Re\, k(t)$ <br> 1) $Re\, k(\tau_s^{IBM}) = 6 \times 10^{-3}$ <br> 2) $Re\, k(\tau_s^{IonQ}) = 1 \times 10^{-3}$ | $Re\, k(t)$ | 0 |

From the above table, the decoherence function $Re\, k(t)$ at switching time $\tau_s$ can be estimated for each NISQ machine as, $Re\, k(\tau_s^{IBM}) = 6 \times 10^{-3}$, and $Re\, k(\tau_s^{IonQ}) = 1 \times 10^{-3}$, respectively. It is interesting to note that the $Re\, k(t)$ can be approximated as

$$Re\, k(t) \approx \frac{2}{\pi} \Gamma_0 \omega_c \tau_s \left[ \frac{t}{\tau_s} + \frac{1}{2} \left( \frac{t}{\tau_s} \right)^2 \right]. \tag{32}$$

The change of $Re\, k(t)$ vs $\frac{t}{\tau_s}$ for various $\Gamma_0 \omega_c \tau_s$ is shown in figure 2. This figure shows that $Re\, k(t)$ increases almost quadratic for $\frac{t}{\tau_s}$.

Then, from Eqs. (31) and (32), we obtain

$$\begin{aligned}\rho_{11}^c(t) &\approx \left( 1 - \frac{4}{\pi} \Gamma_0 \omega_c \tau_s \left[ \frac{t}{\tau_s} + \frac{1}{2} \left( \frac{t}{\tau_s} \right)^2 \right] \right) \rho_{11}^m(0) \\ &= exp\left[ \frac{2}{\pi} \Gamma_0 \omega_c \tau_s \right] exp\left[ -\frac{2}{\pi} \Gamma_0 \omega_c \tau_s \left( \frac{t}{\tau_s} + 1 \right)^2 \right] \rho_{11}^m(0) ,\end{aligned} \tag{33}$$

with the evolution process is governed by Gaussian in time and the peak value of the density matrix element occurs at $t=0$ is unity as depicted in figure 4. The NISQ system parameter $\Gamma_0 \omega_c \tau_s$ is $2.548 \times 10^{-3}$ for ibm-guadrupe and $4.26 \times 10^{-4}$ for IonQ, respectively. Our non-Markovian model allows bit-flip as a noise source.

We now consider the non-Markovian error associated with CNOT gate operation depicted in figure 5. For CNOT gate operation, the multiplet basis is given by

$$\begin{aligned} |1>^{CNOT} &= |00>, \\ |2>^{CNOT} &= |01>, \\ |3>^{CNOT} &= \frac{1}{\sqrt{2}} (|10> + |11>) , \\ |4>^{CNOT} &= \frac{1}{\sqrt{2}} (|10> - |11>) . \end{aligned} \tag{34}$$

The noisy evolution operator $V(t)$ represented by the multiplet basis of Eq. (35) is given by

$$V(t) = \begin{bmatrix} 1-2Re\,k(t) & Re\,k(t) & \frac{Re\,k(t)}{2} & \frac{Re\,k(t)}{2} \\ Re\,k(t) & 1-2\,Re\,k(t) & \frac{Re\,k(t)}{2} & \frac{Re\,k(t)}{2} \\ \frac{Re\,k(t)}{2} & \frac{Re\,k(t)}{2} & 1-\frac{3Re\,k(t)}{2} & \frac{Re\,k(t)}{2} \\ \frac{Re\,k(t)}{2} & \frac{Re\,k(t)}{2} & \frac{Re\,k(t)}{2} & 1-\frac{3Re\,k(t)}{2} \end{bmatrix}, \quad (35)$$

from Eqs. (67), (69), (71) ad (73) of the Supplementary Information.

Then the QEM recovery operator $R_{QEM}(t)$ is obtained from Eqs. (25) and (35) and is given by

$$R_{QEM}(t) = \begin{bmatrix} C & D & B & B \\ D & C & B & B \\ B & B & E & F \\ B & B & F & E \end{bmatrix}, \quad (36)$$

where

$$\begin{aligned} B &= \frac{-0.5\alpha + 2.125\alpha^2 - 1.875\alpha^3}{1 - 5.5\alpha + 8.3125\alpha^2 - 1.5\alpha^3 - 2.8125\alpha^4}, \\ C &= \frac{1 - 3.5\alpha + 2.8125\alpha^2}{1 - 5.5\alpha + 8.3125\alpha^2 - 1.5\alpha^3 - 2.8125\alpha^4}, \\ D &= \frac{-\alpha + 2\alpha^2 - 0.9375\alpha^3}{1 - 5.5\alpha + 8.3125\alpha^2 - 1.5\alpha^3 - 2.8125\alpha^4}, \\ E &= \frac{1 - 4.75\alpha + 5.5\alpha^2 - 0.75\alpha^3}{1 - 5.5\alpha + 8.3125\alpha^2 - 1.5\alpha^3 - 2.8125\alpha^4}, \\ F &= \frac{-0.5\alpha + 2.5\alpha^2 - 3\alpha^3}{1 - 5.5\alpha + 8.3125\alpha^2 - 1.5\alpha^3 - 2.8125\alpha^4} \end{aligned} \quad (37)$$

and $\alpha = Re\,k(t)$.

The QEM recovery operator can be expanded by 16 Dirac matrices $E_{ij} = (\sigma_i \otimes I)(I \otimes \sigma_j)$ or equivalently, $\gamma_r = I,\, \gamma_\mu,\, \gamma_\mu \gamma_\nu (\mu < \nu),\, \gamma_5 \gamma_\mu,\, \gamma_5 = \gamma_0 \gamma_1 \gamma_2 \gamma_3$ for $\mu = 0, 1, 2, 3$ [28]. Here

$$\gamma_i = \begin{pmatrix} 0 & \sigma_i \\ \sigma_i & 0 \end{pmatrix} (i = 1, 2, 3) \text{ and } \gamma_0 = i\begin{pmatrix} I & 0 \\ 0 & -I \end{pmatrix}.$$

After some mathematical manipulations, we expand the QEM recovery operator as

$$R_{QEM}(t) = \frac{C+E}{2}I + D(I \otimes \sigma_1) + \frac{F+2B-D}{2}(\sigma_1 \otimes I) - \frac{E-C}{2}(\sigma_3 \otimes I) \\ + B(\sigma_1 \otimes I)(I \otimes \sigma_1) - \frac{F-D}{2}(\sigma_3 \otimes \sigma_1). \quad (38)$$

From Eq. (26), we obtain the QEM cost function as

$$c(t) = \frac{|C+E|}{2} + \frac{|C-E|}{2} + 2|B| + |D| + |F-D|$$

$$= \frac{|2 - 8.25\alpha + 8.3125\alpha^2 - 0.75\alpha^3|}{2|1 - 5.5\alpha + 8.3125\alpha^2 - 1.5\alpha^3 - 2.8125\alpha^4|} + \frac{|1.25\alpha - 2.6875\alpha^2 + 0.75\alpha^3|}{2|1 - 5.5\alpha + 8.3125\alpha^2 - 1.5\alpha^3 - 2.8125\alpha^4|}$$

$$+ 2\frac{|-0.5\alpha + 2.125\alpha^2 - 1.875\alpha^3|}{|1 - 5.5\alpha + 8.3125\alpha^2 - 1.5\alpha^3 - 2.8125\alpha^4|} + \frac{|-\alpha + 2\alpha^2 - 0.9375\alpha^3|}{|1 - 5.5\alpha + 8.3125\alpha^2 - 1.5\alpha^3 - 2.8125\alpha^4|}$$

$$+ \frac{|0.5\alpha + 0.5\alpha^2 - 2.0625\alpha^3|}{|1 - 5.5\alpha + 8.3125\alpha^2 - 1.5\alpha^3 - 2.8125\alpha^4|}, \quad \alpha = Re\, k(t). \tag{39}$$

Figure 6 presents a graphical representation of the quantum error mitigation cost function with respect to the normalized gate operation time $\frac{t}{\tau_s}$ for varying coupling strengths $\Gamma_0 \omega_c \tau_s$ between a quantum system and its environment. The environment is modeled using a set of simple harmonic oscillators, which serve as a fundamental approximation for the intricate interactions occurring between the quantum system and its surroundings. The quantum error mitigation cost function is a quantitative measure of the deviation or error between the expected output and the actual outcome of a quantum computation. It helps assess the effectiveness of error reduction techniques applied to a quantum computing system. The normalized gate operation time $\frac{t}{\tau_s}$ is a dimensionless parameter that compares the actual gate operation time (t) to a characteristic time scale ($\tau_s$) of the system. This normalization enables the comparison of quantum error mitigation strategies across various systems and gate operation times.

In Figure 6, the coupling strength signifies the intensity of interaction between the quantum system and the environment. A higher coupling strength corresponds to a stronger connection, which can result in a greater susceptibility to non-Markovian noise and other environmental influences. The simple harmonic oscillators offer a convenient method to model these interactions and their impact on the quantum system's behavior. The plot in Figure 6 reveals that the cost function increases as the coupling strength becomes larger. This observation indicates that quantum systems with more substantial interactions with their environment are more prone to errors and necessitate more effective and efficient error mitigation strategies to ensure reliable performance. In conclusion, Figure 6 highlights the importance of accounting for the coupling strength between the quantum system and its environment when developing and evaluating quantum error mitigation techniques. By comprehending how the intensity of these interactions influences the cost function and the effectiveness of error mitigation strategies, researchers can design customized approaches to improve the performance and reliability of quantum computing systems amidst various environmental disturbances.

**Discussion**

In this study, we aim to quantify the errors in individual logical gates due to non-Markovian noise in quantum circuits, as well as determine the cost function for quantum error mitigation. The results of our investigation demonstrate a strong correlation between the theoretical predictions and the experimental data obtained from an ion-trap based Noisy Intermediate-Scale Quantum (NISQ) device. We discovered that the non-Markovian approach provides a reasonable description of the fluctuations in the output quantum states for both identity and CNOT gate operations. These findings suggest that the non-Markovian approach serves as a valuable tool for analyzing the impact of errors and decoherence on the output quantum states of NISQ devices. As the coupling strength increases, the cost function for quantum error mitigation also rises.

The ideal operation of a circuit can be represented as a sequence of ideal logic gates $\{\widehat{U}_n\}$, with an expected outcome of $\ll \widehat{O}|\rho_{out} \gg\, =\, \ll \widehat{O}| \prod_{m=1}^{n} \widehat{U}_m |\rho_{in} \gg \{\widehat{U}_n\}$ [7]. While modeling noise in a quantum system can be difficult, an alternative approach using a correlated Pauli-Lindblad super-operator $\widehat{L}(\rho) = \sum_k \lambda_k \left( P_k \rho P_k - \rho \right)$ for a set of Pauli matrices $P_k$ has been proposed [26].

Our theoretical model for non-Markovian errors, based on the Caldeira-Leggett model, demonstrates a higher degree of agreement with the IonQ machine for the CNOT operation. This is likely due to the fluctuations of the qubit state in an ion trap, which can be better described by the Caldeira-Leggett interaction model.

In conclusion, this paper presents a non-Markovian approach to analyzing quantum state fluctuations in NISQ devices that interact with their environment, which is modeled by simple harmonic oscillators as a noise source. We also provide a quantitative model of a cost function for quantum error mitigation. To achieve this, we employed a projection operator method and advanced and retarded propagators in time. The formalism can be applied to model any quantum system, given specific forms of the system, reservoir, and interaction Hamiltonians.

The authors derived an analytical form of the reduced-density-operator for the output quantum states in a time-convolutionless form and compared the results with experimental data from ion-trap and superconducting quantum computing systems. The results exhibit a strong agreement between the theory and experimental outcomes. The error model and the cost function, derived from the time-convolutionless equation and incorporating non-Markovian effects, can serve as foundational elements for quantum error mitigation in noisy quantum circuits.

**References**

1. Unruh, W. G. Maintaining coherence in quantum computers. *Phys. Rev. A* **51**, 992 (1995).

2. Bauer, B., Wecker, D., Millis, A. J., Hastings, M. B. & Troyer, M. Hybrid quantum-classical approach to correlated materials. *Phys. Rev. X* **6**, 031045 (2016).

3. Li, Y. & Benjamin, S. C. Efficient variational quantum simulator incorporating active error minimization. *Phys. Rev. X* **7**, 021050 (2017).

4. McClean, J. R., Kimchi-Schwartz, M. E., Carter, J. & De Jong, W. A. Hybrid quantum-classical hierarchy for mitigation of decoherence and determination of excited states. *Phys. Rev. A* **95**, 042308 (2017).

5. Colless, J. I. et al. Computation of molecular spectra on a quantum processor with an error-resilient algorithm. *Phys. Rev. X* **8**, 011021 (2018).

6. Temme, K., Bravyi, S. & Gambetta, J. M. Error mitigation for short-depth quantum circuits. *Phys. Rev. Lett.* **119**, 180509 (2017).

7. Endo, S., Benjamin, S. C. & Li, Y. Practical quantum error mitigation for near-future applications. *Phys. Rev. X* **8**, 031027 (2018).

8. Cai, Z. et al. Quantum error mitigation. Preprint at https://arxiv.org/abs/2210.00921 (2022).

9. van Vliet, K. M. Markov approach to density fluctuations due to transport and scattering. I. Mathematical formalism. *J. Math. Phys.* **12**, 1981-1998 (1971).

10. Hakoshima, H., Matsuzaki, Y. & Endo, S. Relationship between costs for quantum error mitigation and non-Markovian measures. Phys. Rev. A 103, 012611 (2021).

11. Hall, M. J. W. Canoniacal form of master equations and characterization of non-Markovianity. Phys. Rev. A 89, 042120 (2014).

12. De Vega, I. & Alonso, D. Dynamics of non-Markovian quantum systems. Rev. Mod. Phys. 89, 015001 (2017).

13. Leggett, A. J. et al. Dynamics of the dissipative two-state system. *Rev. Mod. Phys.* **59**, 1 (1987).

14. Amin, M. H. S., Love, P. J. & Truncik, C. J. S. Thermally assisted adiabatic quantum computation. *Phys. Rev. Lett.* **100**, 060503 (2008).

15. Amin, M. H. S. & Brito, F. Non-Markovian incoherent quantum dynamics of a two-state system. *Phys. Rev. B* **80**, 214302 (2009).

16. Johnson, M. W. et al. Quantum annealing with manufactured spins. *Nature* **473**, 194-198 (2011).

17. Farhi, E. et al. A quantum adiabatic evolution algorithm applied to random instances of an NP-complete problem. *Science* **292**, 472-475 (2001).

18. Ahn, D., Oh, J. H., Kimm, K. & Hwang, S. W. Time-convolutionless reduced-density-

operator theory of a noisy quantum channel: Two-bit quantum gate for quantum-information processing. *Phys. Rev. A* **61**, 052310 (2000).

19. Ahn, D., Lee, J., Kim, M. S. & Hwang, S. W. Self-consistent non-Markovian theory of a quantum-state evolution for quantum-information processing. *Phys. Rev. A* **66**, 012302 (2002).

20. Ahn, D. Time-convolutionless reduced-density-operator theory of an arbitrary driven system coupled to a stochastic reservoir: Quantum kinetic equations for semiconductors. *Phys. Rev. B* **50**, 8310 (1994).

21. Tokuyama, M. & Mori, H. Statistical-mechanical theory of random frequency modulations and generalized Brownian motions. *Prog. Theor. Phys.* **55**, 411-429 (1976).

22. Hashitsumae, N., Shibata, F. & Shingu, M. Quantal master equation valid for any time scale. *J. Stat. Phys.* **17**, 155-169 (1977).

23. Saeki, M. A generalized equation of motion for the linear response. I: General theory. *Prog. Theor. Phys.* **67**, 1313-1331 (1982).

24. Saeki, M. Generalized master equations for driven system. *J. Phys. Soc. Jpn.* **55**, 1846-1860 (1986).

25. Loss, D. & DiVincenzo, D. P. Quantum computation with quantum dots. *Phys. Rev. A* **57**, 120 (1998).

26. van den Berg, E., Minev, Z. K. & Temme, K. Model-free readout-error mitigation for quantum expectation values. *Phys. Rev. A* **105**, 032620 (2022).

27. Gradshteyn, I. S. & Ryzhik, I. M. Table of integrals, series, and products. (Academic Press, 2007).

28. Jauch J. M. & Rohrlich, F. The theory of photons and elctrons (Springer-Verlag, New York, 1976)

## Acknowledgments

The authors thank Dr. Hansol Noh for his valuable comments. This work was supported by Korea National Research Foundation (NRF) grant No. NRF-2023R1A2C1003570, ICT R&D program of IITP-2023- 2021-0-01810, RS-2023-00225385, AFOSR grant FA2386-21-1-0089, AFOSR grant FA2386-22-1-4052, and Amazon Web Services. We also acknowledge the use of IBM Quantum services for this work. The views expressed are those of the authors, and do not reflect the official policy or position of IBM or the IBM Quantum team.

## Author contributions

D.A. developed the main idea, carried out the analysis, and and wrote the main manuscript. B. P. carried out the experiment. All authors reviewed the manuscript.

## Data availability

The data generated during the current study are available from the corresponding author on reasonable request.

## Competing interests

The authors declare no competing interests.

**Figure legends**

**Figure 1.** Schematic representation of the open quantum system consisting of a two-state quantum subsystem $H_S$, an environment $H_B$, and their mutual interaction $H_{int}$. The quantum subsystem $H_S$ represents the qubits within the quantum device, while the environment $H_B$ models the external influences that contribute to noise and decoherence. The interaction $H_{int}$ between the quantum subsystem and the environment plays a crucial role in determining the system's dynamics and the noise sources affecting its performance. This comprehensive view of the open quantum system provides a framework for understanding and analyzing error sources and devising effective error mitigation strategies.

**Figure 2.** Schematic representation of the identity operation circuit. All qubits are initialized to the state |0⟩ before the circuit execution. The circuit was implemented on two distinct quantum computing devices, ibm_guadalupe and IonQ, with each device running the circuit 1000 times to ensure reliable comparative analysis of the outcomes.

**Figure 3.** Plot of $Re\, k(t)$ vs $\frac{t}{\tau_s}$, where $\tau_s$ is the switching time for the parameters $\Gamma_0 \omega_c \tau_s$ in the range $7.0 \times 10^{-4} \leqslant \Gamma_0 \omega_c \tau_s \leq 7.0 \times 10^{-3}$.

**Figure 4.** Plot of probability distribution for $\rho_{11}(t)$ given in Eq. (42) vs $\frac{t}{\tau_s}$, where $\tau_s$ is the switching time for the parameters $\Gamma_0 \omega_c \tau_s$ in the range $7.0 \times 10^{-4} \leqslant \Gamma_0 \omega_c \tau_s \leq 7.0 \times 10^{-3}$.

**Figure 5.** Schematic representation of the circuit for implementing a CNOT gate on the multiplet basis described in Eq. (34). All qubits are initialized to the state |0⟩ before the circuit execution. Several single-qubit gates are applied before the CNOT gate to evolve the initial state |00⟩ to the multiplet basis. These circuits were implemented on both on ibm_guadalupe and IonQ devices 1,000 times each for reliable comparative analysis of the outcomes. (a) CNOT gate implementation on the first basis, |00⟩. (b) CNOT gate implementation on the second basis, ||01⟩.. (c) CNOT gate implementation on the third basis, (|10⟩ + |11⟩)/√2. (d) CNOT gate implementation on the fourth basis, (|10⟩ − |11⟩)/√2.

**Figure 6**. A graphical representation of the quantum error mitigation cost function with respect to the normalized gate operation time $\frac{t}{\tau_s}$ for varying coupling strengths $\Gamma_0 \omega_c \tau_s$ between a quantum system and its environment.

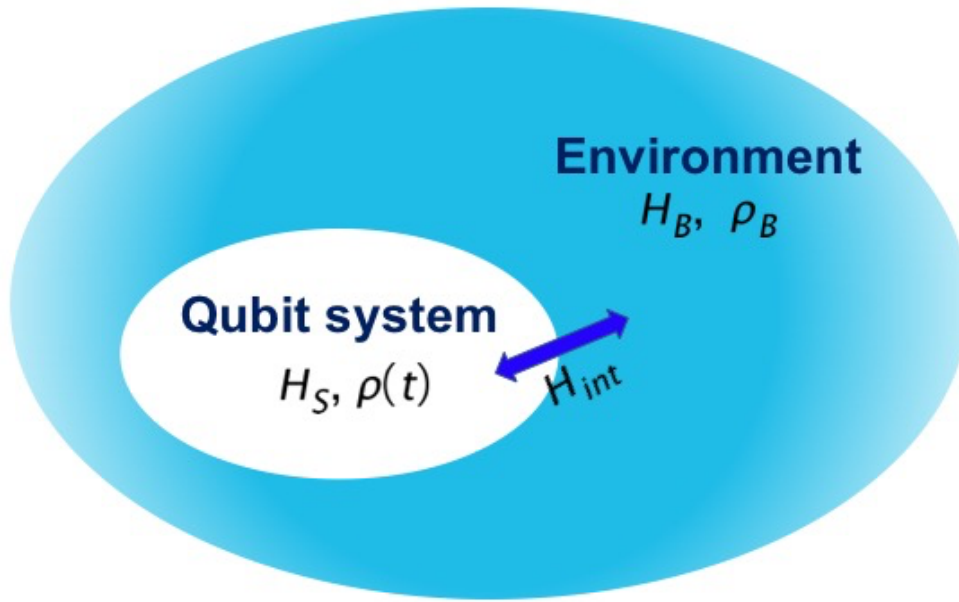

**Fig. 1**

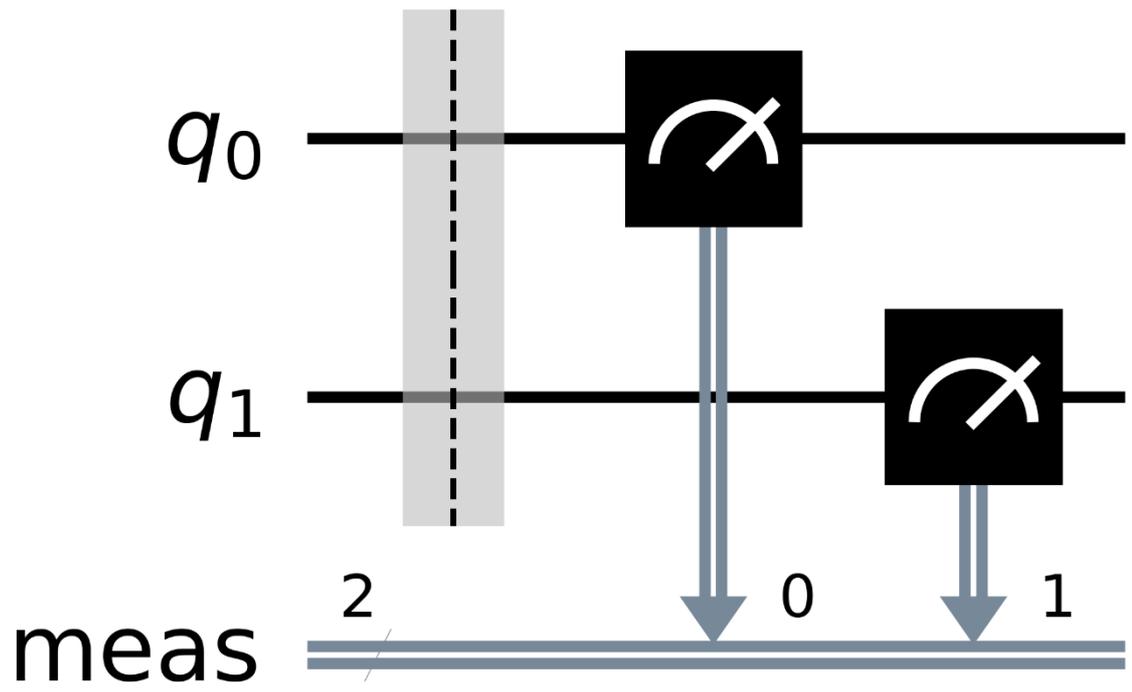

**Fig. 2**

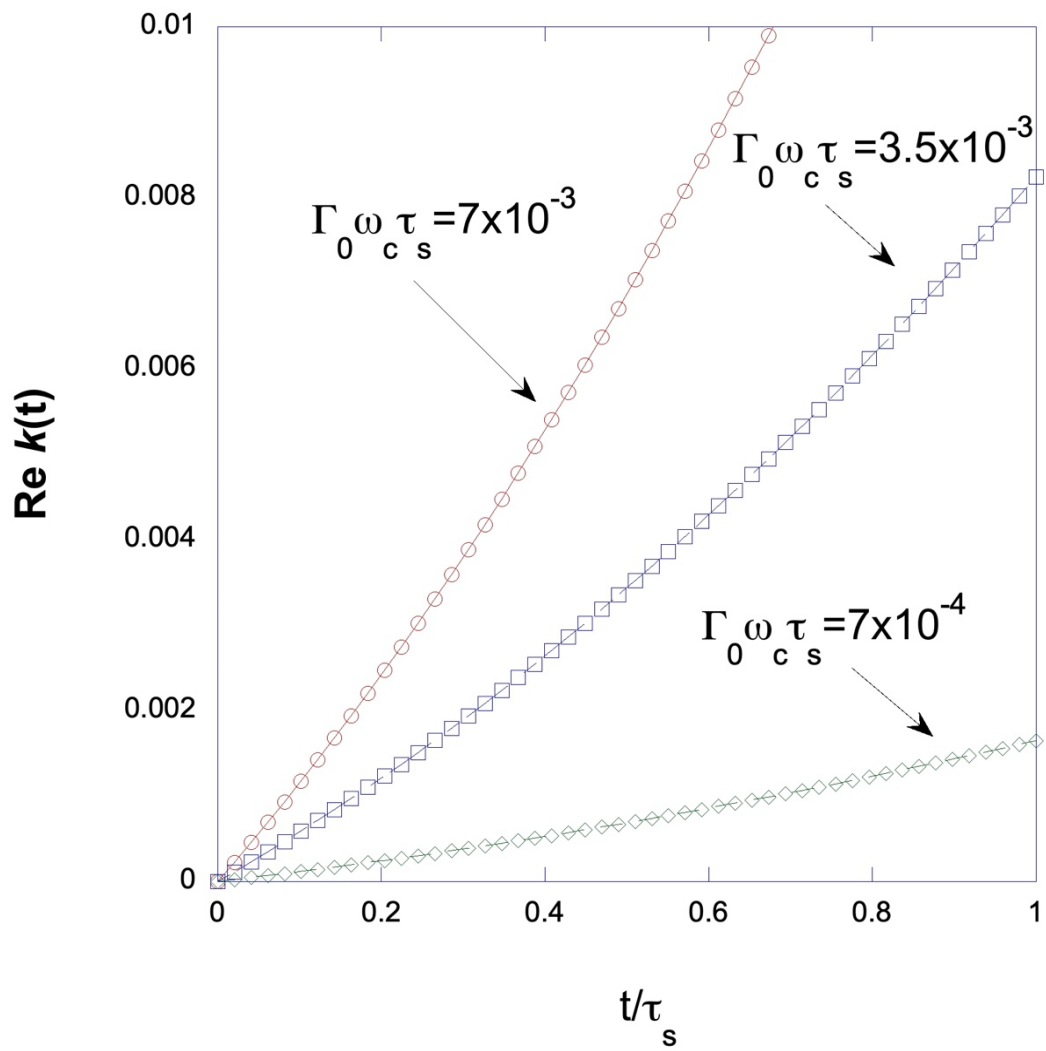

**Fig. 3**

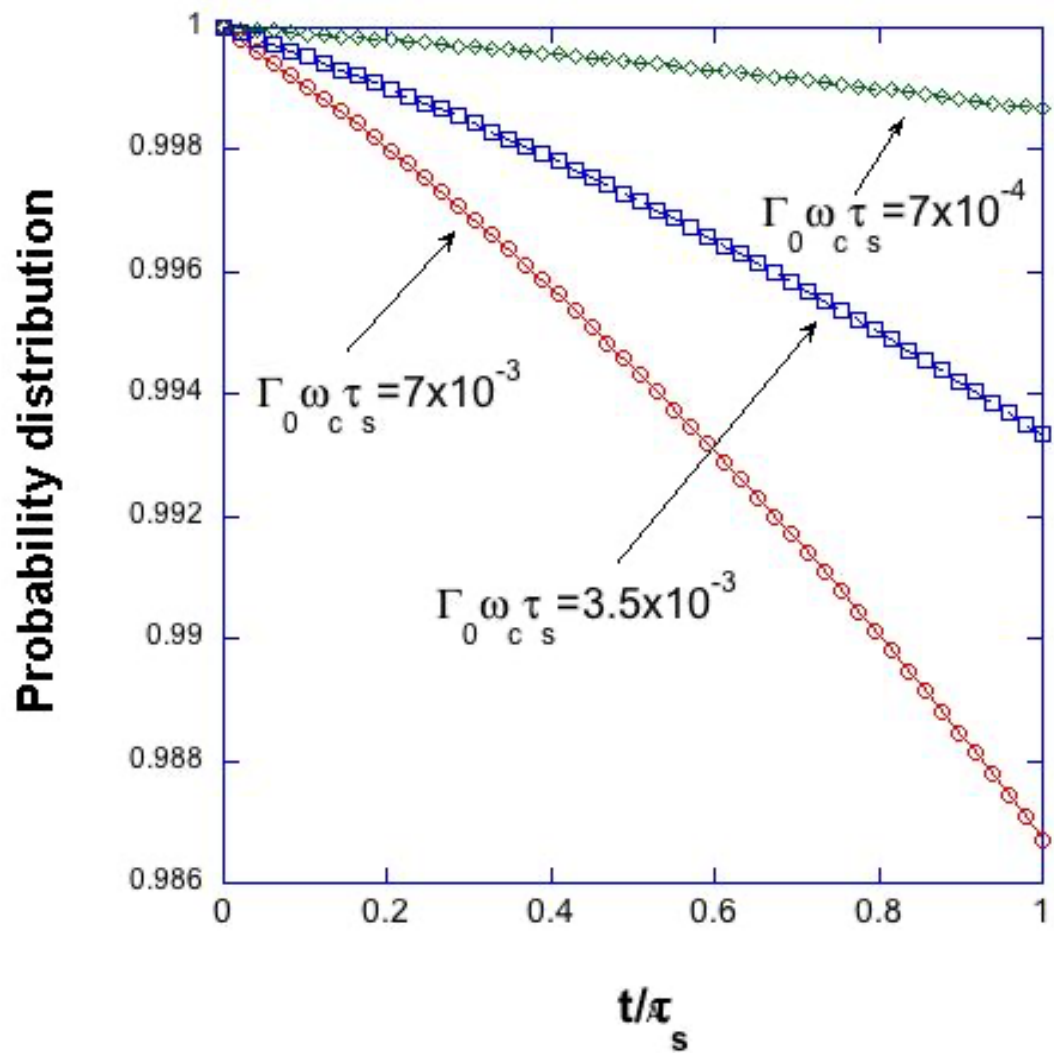

**Fig. 4**

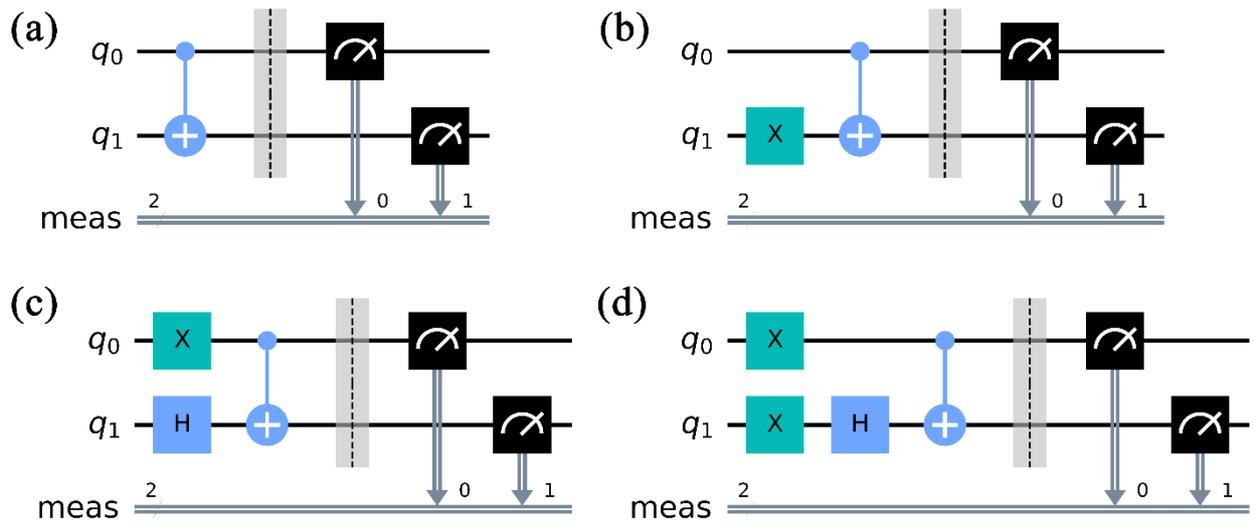

**Fig. 5**

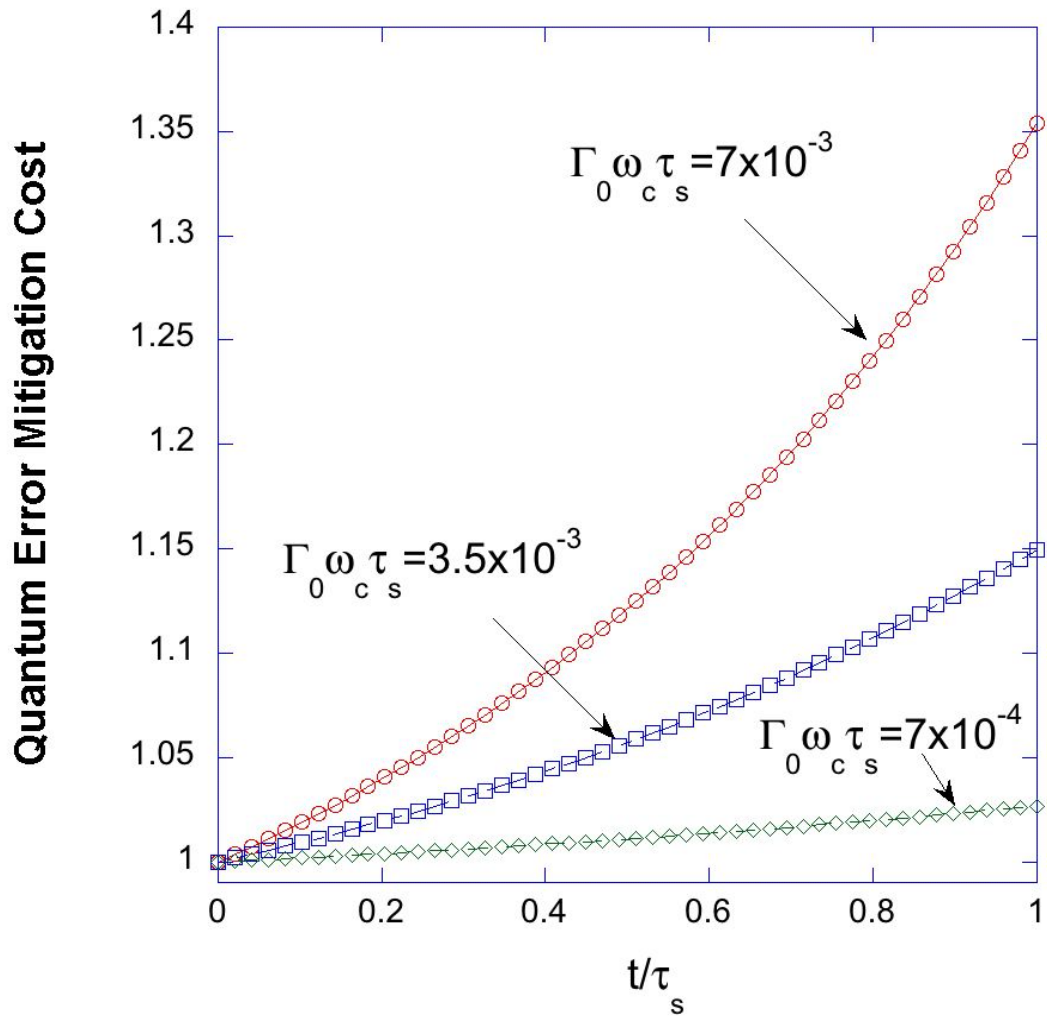

**Fig. 6**

# Supplementary Information for
# Non-Markovian noise sources for quantum error mitigation

Doyeol Ahn[1,2,*] and Byeongyong Park[1,3]

[1]Department of Electrical and Computer Engineering,
University of Seoul, 163 Seoulsiripdae-ro, Tongdaimoon-gu, Seoul 02504, Republic of Korea

[2]First Quantum, Inc, #210, 180 Bangbae-ro, Seocho-gu, Seoul 06586, Republic of Korea

*Corresponding author: dahn@uos.ac.kr

**I. Time-convolutionless evolution of a non-Markovian reduced-density operator**

The total Hamiltonian for an open two-state system is given by [1-5]

$$\widehat{H}_T = \widehat{H}_S(t) + \widehat{H}_B + \widehat{H}_{int} \tag{1}$$

where $\widehat{H}_S(t)$ is the system Hamiltonian for a two-state system, $\widehat{H}_B$ the Hamiltonian acting on the reservoir or an environment, $\widehat{H}_{int}$ is the interaction between the system and the environment (as depicted in figure 1).

The equation of motion for the total density operator $\rho_T(t)$ of the total system is given by a quantum Liouville equation [1]

$$\frac{d}{dt}\rho_T(t) = -i[\widehat{H}_T(t), \rho_T(t)] = -i\widehat{L}_T(t)\rho_T(t), \tag{2}$$

where

$$\widehat{L}_T(t) = \widehat{L}_S(t) + \widehat{L}_B(t) + \widehat{L}_{int}(t) \tag{3}$$

is the Liouville super operator in one-to-one correspondence with the Hamiltonian. Here, we use the unit in which $\hbar = 1$. In order to derive an equation and to solve for a system alone, it is convenient to use the projection operators which decompose the total system by eliminating the degrees of freedom for the reservoir. We define thine-independent projection operator $\underline{P}$ and $\underline{Q}$ given by [1,2]

$$\underline{P}X = \rho_B tr_B X, \underline{Q} = 1 - \underline{P} \tag{4}$$

for any dynamical variable $X$. Here $tr_B$ denotes a partial trace over the quantum reservoir and $\rho_B$ is the density matrix of the reservoir. The projection operators satisfy the following operator identities:

$$\underline{P}^2 = \underline{P}, \ \underline{Q}^2 = \underline{Q} \ \underline{P} \ \underline{Q} = \underline{Q}\underline{P} = 0, \tag{5a}$$

$$\underline{P}\widehat{L}_S(t) = \widehat{L}_S(t)\underline{P}, \ \underline{P}\widehat{L}_B = \widehat{L}_B\underline{P} = 0, \ \underline{P}\widehat{L}_{int}\underline{P} = 0, \tag{5b}$$

and $\quad \underline{Q}\widehat{L}_S(t) = \widehat{L}_S(t)\underline{Q}, \ \underline{Q}\widehat{L}_B = \widehat{L}_B\underline{Q} = \widehat{L}_B. \tag{5c}$

The information of the system is then contained in the reduced density operator

$$\rho(t) = tr_B \rho_T(t) = tr_B \underline{P} \rho_T(t). \tag{6}$$

If we multiply equation (2) by $\underline{P}$ and $\underline{Q}$ on the left side, we obtain coupled equations for $\underline{P}\rho_T(t)$ and $\underline{Q}\rho_T(t)$ as follows:

$$\frac{d}{dt}\underline{P}\rho_T(t) = -i\underline{P}\hat{L}_T(t)\underline{P}\rho_T(t) - -i\underline{P}\hat{L}_T(t)\underline{Q}\rho_T(t), \tag{7a}$$

and

$$\frac{d}{dt}\underline{Q}\rho_T(t) = -i\underline{Q}\hat{L}_T(t)\underline{Q}\rho_T(t) - -i\underline{Q}\hat{L}_T(t)\underline{P}\rho_T(t), \tag{7b}$$

where we have modified the Eq. (2) as

$$\frac{d}{dt}\rho_T(t) = -i\hat{L}_T(t)\rho_T(t) = --i\hat{L}_T(t)\left(\underline{P}+\underline{Q}\right)\rho_T(t).$$

We assume that the system was turned on at $t=0$ and the input state prepared at $t=0$ was isolated with the reservoir such that $\underline{Q}\rho_T(0)=0$. The formal solution of (7b) is given by

$$\begin{aligned}\underline{Q}\rho_T(t) &= -i\int_0^t \underline{H}(t,\tau)\,\underline{Q}\hat{L}_T(\tau)\,\underline{P}\rho_T(\tau)\,d\tau + \underline{H}(t,0)\,\underline{Q}\rho_T(0) \\ &= -i\int_0^t \underline{H}(t,\tau)\,\underline{Q}\hat{L}_T(\tau)\,\underline{P}\rho_T(\tau)\,d\tau\,.\end{aligned} \tag{8}$$

where the projected propagator $\underline{H}(t,\tau)$ of the total system is defined by

$$\underline{H}(t,\tau) = \underline{T}exp\left[-i\int_\tau^t \underline{Q}\hat{L}_T(s)\,\underline{Q}ds\right]. \tag{9}$$

Here $\underline{T}$ is the time-ordering operator. We also introduce an anti-time evolution operator $\underline{G}(t,\tau)$ which is defined by

$$\underline{G}(t,\tau) = \underline{T}^C exp\left[i\int_\tau^t \hat{L}_T(s)\,ds\right], \tag{10}$$

such that $\rho_T(\tau) = \underline{G}(t,\tau)\rho_T(t)$. \hfill (11)

It is not very difficult to show [1] that

$$\underline{G}(t,\tau)\,\underline{G}(s,t) = \underline{G}(s,\tau),\ \underline{H}(t,\tau)\,\underline{H}(\tau,s) = \underline{H}(t,s). \tag{12}$$

Here $\underline{T}^C$ is anti-time-ordering operator.

From Eqs. (8)-(11), we obtain

$$\underline{Q}\rho_T(t) = -i\int_0^t \underline{H}(t,\tau)\,\underline{Q}\hat{L}_T(\tau)\,\underline{P}\,\underline{G}(t,\tau)\left(\underline{P}+\underline{Q}\right)\rho_T(t) \tag{13}$$

which is obviously in time-convolutionless form. Eq. (13) can be rewritten as

$$\begin{aligned}\left[1 + i\int_0^t \underline{H}(t,\tau)\,\underline{Q}\,\hat{L}_T(\tau)\,\underline{P}\,\underline{G}(t,\tau)\,d\tau\right]\underline{Q}\rho_T(t) \\ = -i\int_0^t d\tau\underline{H}(t,\tau)\,\underline{Q}\,\hat{L}_T(\tau)\,\underline{P}\,\underline{G}(t,\tau)\,\underline{P}\,\rho_T(t)\,.\end{aligned} \tag{14}$$

We introduce new super operators $\hat{g}(t)$ and $\hat{\theta}(t)$ which are given by

$$\hat{g}(t) = 1 + i\int_0^t d\tau \underline{H}(t,\tau) \underline{Q}\hat{L}_T(\tau) \underline{P}\ \underline{G}(t,\tau) = \hat{\theta}^{-1}(t). \tag{15}$$

Then from Eqs. (13)-(15), we obtain

$$\underline{Q}\rho_T(t) = \{\hat{\theta}(t) - 1\}\underline{P}\rho_T(t) \tag{16}$$

and

$$\frac{d}{dt}\underline{P}\rho_T(t) + i\underline{P}\hat{L}_T(t)\underline{P}\rho_T(t) = -i\underline{P}\hat{L}_T(t)\{\hat{\theta}(t) - 1\}\underline{P}\rho_T(t). \tag{17}$$

After some mathematical manipulations, we obtain the formal solution of Eq. (17), which is given by

$$\begin{aligned}
\underline{P}\rho_T(t) &= \widehat{U}(t,0)\underline{P}\rho_T(0) - i\int_0^t ds\,\widehat{U}(t,s)\underline{P}\hat{L}_T(s)\{\hat{\theta}(s) - 1\}\underline{P}\rho_T(s) \\
&= \widehat{U}(t,0)\underline{P}\rho_T(0) - i\int_0^t ds\,\widehat{U}(t,s)\underline{P}\hat{L}_T(s)\{\hat{\theta}(s) - 1\}\underline{P}\ \underline{G}(t,s)\rho_T(t) \\
&= \widehat{U}(t,0)\underline{P}\rho_T(0) - i\int_0^t ds\,\widehat{U}(t,s)\underline{P}\hat{L}_T(s)\{\hat{\theta}(s) - 1\}\underline{P}\ \underline{G}(t,s)\underline{P}\rho_T(t) \\
&\quad - i\int_0^t ds\,\widehat{U}(t,s)\underline{P}\hat{L}_T(s)\{\hat{\theta}(s) - 1\}\underline{P}\ \underline{G}(t,s)\underline{Q}\rho_T(t) \\
&= \widehat{U}(t,0)\underline{P}\rho_T(0) - i\int_0^t ds\,\widehat{U}(t,s)\underline{P}\hat{L}_T(s)\{\hat{\theta}(s) - 1\}\underline{P}\ \underline{G}(t,s)\underline{P}\rho_T(t) \\
&\quad - i\int_0^t ds\,\widehat{U}(t,s)\underline{P}\hat{L}_T(s)\{\hat{\theta}(s) - 1\}\underline{P}\ \underline{G}(t,s)\{\hat{\theta}(t) - 1\}\underline{P}\rho_T(t) \\
&= \widehat{U}(t,0)\underline{P}\rho_T(0) - i\int_0^t ds\,\widehat{U}(t,s)\underline{P}\hat{L}_T(s)\{\hat{\theta}(s) - 1\}\underline{P}\ \underline{G}(t,s)\hat{\theta}(t)\underline{P}\rho_T(t),
\end{aligned} \tag{18}$$

where

$$\widehat{U}(t,\tau) = \underline{T}exp\left[-i\int_\tau^t ds\ \underline{P}\hat{L}_T(s)\underline{P}\right] \tag{19}$$

is the projected propagator of the total system.

From, $\underline{P}\hat{L}_T(t)\underline{P} = \underline{P}\left(\hat{L}_S(t) + \hat{L}_B + \hat{L}_{int}\right)\underline{P} = \underline{P}\ \hat{L}_S(t)\underline{P} = \hat{L}_S(t)\underline{P}^2 = \hat{L}_S(t)\underline{P}$ , we obtain

$$\begin{aligned}
\widehat{U}(t,s)\underline{P} &= \underline{T}exp\left[-i\int_s^t d\tau \underline{P}\hat{L}_T(\tau)\underline{P}\right]\underline{P} \\
&= \underline{T}\,exp\left[-i\int_s^t d\tau \hat{L}_S(\tau)\underline{P}\right]\underline{P} \\
&= \underline{T}\,exp\left[-i\int_s^t d\tau \hat{L}_S(\tau)\right]\underline{P} = \widehat{U}_S(t,s)\underline{P}.
\end{aligned} \tag{20}$$

Here $\widehat{U}_S(t,s)$ is the propagator of the system.

From Eqs. (4) and (6), we get

$$\widehat{U}(t,s)\,\underline{P}\widehat{L}_T(s)\,\{\widehat{\theta}(s)-1\}\underline{P}\,\underline{G}(t,s)\,\widehat{\theta}(t)\,\underline{P}\rho_T(t)$$
$$=\widehat{U}_S(t,s)\,\rho_B\,tr_B\{\underline{P}\widehat{L}_T(s)\,\{\widehat{\theta}(s)-1\}\rho_B\}tr_B\{\underline{G}(t,s)\widehat{\theta}(t)\rho_B\}\rho(t)$$

By the way, by substituting Eq. (7) into Eqs. (18)-(20), we obtain

$$\rho_B\rho(t)=\rho_B\widehat{U}_S(t,0)\rho(0)-i\rho_B\int_0^t ds\widehat{U}_S(t,s)tr_B[\widehat{L}_T(s)\{\widehat{\theta}(s)-1\}\rho_B]tr_B[\underline{G}(t,s)\widehat{\theta}(t)\rho_B]\rho(t),$$
(21)

or

$$\left(1+i\int_0^t ds\widehat{U}_S(t,s)tr_B[\widehat{L}_T(s)\{\widehat{\theta}(s)-1\}\rho_B]tr_B[\underline{G}(t,s)\widehat{\theta}(t)\rho_B]\right)\rho(t)$$
$$=\widehat{U}_S(t,s)\rho(t)\,.$$
(22)

If we define $\widehat{W}(t)$ by

$$\widehat{W}(t)=1+i\int_0^t ds\widehat{U}_S(t,s)\,tr_B[\widehat{L}_T(s)\{\widehat{\theta}(s)-1\}\rho_B]tr_B[\underline{G}(t,s)\widehat{\theta}(t)\rho_B],\qquad(23)$$

Then, the evolution operator for the reduced density operator $\rho(t)$ is given by

$$\rho(t)=\widehat{W}^{-1}(t)\,\widehat{U}_S(t,0)\,\rho(0)=\widehat{V}(t)\rho(0)\,,\qquad(24)$$

where the super-operator $\widehat{V}(t)$ for the evolution of the reduced density operator is defined by $\widehat{V}(t)=\widehat{W}^{-1}(t)\,\widehat{U}_S(t,0)$. Within the Born approximation, we have

$$\widehat{V}^{(2)}(t)=\left\{1-\int_0^t ds\int_0^s d\tau\widehat{U}_S(t,s)\,tr_B[\widehat{L}_{int}\widehat{U}_o(s,\tau)\widehat{L}_{int}\widehat{U}_o^{-1}(s,\tau)\rho_B]\widehat{U}_o^{-1}(t,s)\right\}\widehat{U}_S(t,0)\,,\quad(25)$$

where

$$\widehat{U}_o(t)=\underline{T}exp\left[-i\int_0^t d\tau\left(\widehat{L}_S(\tau)+\widehat{L}_B\right)\right]$$
$$=exp(-it\widehat{L}_B)\underline{T}exp\left[-i\int_0^t d\tau\left(\widehat{L}_S(\tau)\right)\right]\qquad(26)$$
$$=\widehat{U}_B(t)\,\widehat{U}_S(t)\,.$$

A detailed derivation of Eq. (25) is given in the below:

We start with Eq. (23)

$$\widehat{W}(t)=1+i\int_0^t ds\widehat{U}_S(t,s)\,tr_B[\widehat{L}_T(s)\{\widehat{\theta}(s)-1\}\rho_B]tr_B[\underline{G}(t,s)\widehat{\theta}(t)\rho_B].\qquad(23)$$

We define

$$\widehat{\Sigma}(t)=1-\widehat{\theta}(t)$$
$$=-i\int_0^t d\tau\underline{H}(t,\tau)\,\underline{Q}\,\widehat{L}_T(\tau)\,\underline{P}\,\underline{G}(t,\tau)\,,\qquad(27)$$

then

$$\widehat{\theta}^{-1}(t) = 1 - \widehat{\Sigma}(t),$$
$$\widehat{\theta}(t) - 1 = \frac{\widehat{\Sigma}(t)}{1 - \widehat{\Sigma}(t)}. \tag{28}$$

Also, we have
$$\underline{P}\,\widehat{L}_T(t)\{\widehat{\theta}(t) - 1\}$$
$$=\underline{P}\left(\widehat{L}_S(t) + \widehat{L}_B + \widehat{L}_{int}\right)\widehat{\Sigma}(t)\{1 - \widehat{\Sigma}(t)\}^{-1}.$$

The detailed expression for $\widehat{\Sigma}(t)$ becomes

$$\widehat{\Sigma}(t) = -i\int_0^t d\tau \underline{H}(t,\tau)\,\underline{Q}\,\widehat{L}_T(\tau)\,\underline{P}\,\underline{G}(t,\tau)$$
$$= -i\int_0^t d\tau \underline{H}(t,\tau)\,\underline{Q}\,\widehat{L}_{int}(\tau)\,\underline{P}\,\underline{G}(t,\tau) \tag{29}$$
$$= -i\int_0^t d\tau \widehat{U}_o(t)\,\underline{S}(t,\tau)\,\widehat{U}_o^{-1}(\tau)\,\underline{Q}\,\widehat{L}_T(\tau)\,\underline{P}\,\widehat{U}_o(\tau)\,\underline{R}(t,\tau)\,\widehat{U}_o^{-1}(t),$$

where
$$\underline{R}(t,\tau) = \underline{T}^C \exp\left[i\int_\tau^t ds\,\widehat{U}_o^{-1}(s)\,\widehat{L}_{int}\,\widehat{U}_o(s)\right] \tag{30}$$

is the evolution operator in the interaction picture and

$$\underline{S}(t,\tau) = \underline{T}\exp\left[i\int_\tau^t ds\,\underline{Q}\,\widehat{U}_o^{-1}(s)\,\widehat{L}_{int}\,\widehat{U}_o(s)\,\underline{Q}\right] \tag{31}$$

is the projected propagator of the total system in the interaction picture. Then Eq. (23) can be rewritten as

$$\widehat{W}(t) = 1 + i\int_0^t ds\,\widehat{U}_S(t,s)\,tr_B\left[\widehat{L}_{int}\widehat{\Sigma}(s)\{1 - \widehat{\Sigma}(s)\}^{-1}\rho_B\right]tr_B\left[\widehat{U}_o(s)\,\underline{R}(t,s)\,\widehat{U}_o^{-1}\{1 - \widehat{\Sigma}(t)\}^{-1}\rho_B\right].$$
$$\tag{32}$$

where
$$\underline{R}(t,\tau) = \underline{T}^C \exp\left[i\int_\tau^t ds\,\widehat{U}_o^{-1}(s)\,\widehat{L}_{int}\,\widehat{U}_o(s)\right] \tag{33}$$

is the evolution operator in the interaction picture and

$$\underline{S}(t,\tau) = \underline{T}\exp\left[i\int_\tau^t ds\,\underline{Q}\,\widehat{U}_o^{-1}(s)\,\widehat{L}_{int}\,\widehat{U}_o(s)\,\underline{Q}\right] \tag{34}$$

is the projected propagator of the total system in the interaction picture. Then Eq. (23) can be rewritten as

$$\widehat{W}(t) = 1 + i\int_0^t ds\,\widehat{U}_S(t,s)\,tr_B\left[\widehat{L}_{int}\widehat{\Sigma}(s)\{1 - \widehat{\Sigma}(s)\}^{-1}\rho_B\right]tr_B\left[\widehat{U}_o(s)\,\underline{R}(t,s)\,\widehat{U}_o^{-1}\{1 - \widehat{\Sigma}(t)\}^{-1}\rho_B\right].$$
$$\tag{35}$$

Born approximation of Eq. (35) leads to

$$\widehat{W}^{(2)}(t) = 1 + i \int_0^t ds \widehat{U}_S(t,s) \, tr_B\left[\widehat{L}_{int}\widehat{\Sigma}^{(1)}(s)\rho_B\right] tr_B\left[\widehat{U}_o(s)\widehat{U}_o^{-1}\rho_B\right], \quad (36)$$

where

$$\widehat{\Sigma}^{(1)}(s) = -i \int_0^s ds \widehat{U}_o(s) \widehat{U}_o^{-1}(\tau) \underline{Q}\widehat{L}_{int}\underline{P} \, \widehat{U}_o(\tau) \widehat{U}_o^{-1}(s)$$

$$= -i \int_0^s ds \widehat{U}_o(s) \widehat{U}_o^{-1}(\tau) \widehat{L}_{int} \widehat{U}_o(\tau) \widehat{U}_o^{-1}(s) \quad (37)$$

$$= -i \int_0^s ds \widehat{U}_o(s,\tau) \widehat{L}_{int} \widehat{U}_o^{-1}(s,\tau),$$

using the ansatz $\underline{P}\widehat{L}_{int}\underline{P} = 0$. From Eqs. (36) and (37), we get

$$\widehat{W}^{-1(2)}(t) = 1 - i \int_0^t ds \widehat{U}_S(t,s) \, tr_B\left[\widehat{L}_{int}\widehat{\Sigma}^{(1)}(s)\rho_B\right] tr_B\left[\widehat{U}_o(s)\widehat{U}_o^{-1}\rho_B\right]$$

$$= 1 - i \int_0^t ds \widehat{U}_S(t,s) \, tr_B\left[\widehat{L}_{int}\widehat{\Sigma}^{(1)}(s)\rho_B\right] \widehat{U}_S^{-1}(t,s) \quad (38)$$

$$= 1 - \int_0^t ds \int_0^s d\tau \widehat{U}_S(t,s) \, tr_B\left[\widehat{L}_{int}\widehat{U}_o(s,\tau) \widehat{L}_{int}\widehat{U}_o^{-1}(s,\tau)\rho_B\right] \widehat{U}_S^{-1}(t,s).$$

By substituting Eq. (38) into $\widehat{V}^{(2)}(t) = \widehat{W}^{-1(2)}(t) \widehat{U}_S(t,0)$, we obtain Eq. (25). After some mathematical manipulations, $\widehat{V}^{(2)}(t)\rho(0)$ becomes

$$\rho(t)$$
$$= \widehat{V}^{(2)}(t)\rho(0)$$
$$= \widehat{U}_S(t)\rho(0) - \int_0^t ds \int_0^s d\tau \, d\tau \, tr_B\left[\widehat{H}_{int}\widehat{H}_{int}(\tau-s)\rho_B\rho(-s)\right] + \int_0^t ds \int_0^s d\tau \, tr_B\left[\widehat{H}_{int}(\tau-s)\rho_B\rho(-s)\rho_B\widehat{H}_{int}\right]$$
$$+ \int_0^t ds \int_0^s d\tau \, tr_B\left[\widehat{H}_{int}\rho_B\rho(-s)\widehat{H}_{int}(\tau-s)\right] - \int_0^t ds \int_0^s d\tau \, tr_B\left[\rho_B\rho(-s)\widehat{H}_{int}(\tau-s)\widehat{H}_{int}\right].$$

$$(39)$$

Here $\widehat{H}_{int}(\tau-s)$ and $\rho(-s)$ are Heisenberg operators defined by

$$\widehat{H}_{int}(\tau-s) = \exp\left(i\widehat{H}_{int}(\tau-s)\right)\widehat{H}_{int}\exp\left(-i\widehat{H}_{int}(\tau-s)\right),$$
$$\rho(-s) = \exp\left(-i\widehat{H}_S s\right)\rho(0)\exp\left(i\widehat{H}_S s\right),$$

respectively.

**Non-Markovian evolution of two-qubit gate operation**

In this work, we focus on the two-qubit gate operations and model the interaction of the quantum system with the environment during the gate operation by a Caldeira-Leggett model [3-5] where a set of harmonic oscillators are coupled linearly with the system spin by

$$\widehat{H}_{int} = \lambda \sum_{i=1,2, j=1,2,3} \vec{S}_i \cdot \vec{b}_i, \quad S_i^j = \frac{\hbar}{2}\sigma^j \tag{40}$$

where $\sigma_i$ is the Pauli matrices $\sigma_1 = X$, $\sigma_2 = Y$, $\sigma_3 = Z$ and $b_i^j$ is the fluctuating quantum field associated the i$^{th}$ qubit, whose motion is governed by the harmonic-oscillator Hamiltonian. In the evaluation of Eq. (39), we obtain the following relations [1]:

$$tr_B\{b_k^l(t) b_i^j \rho_B\} = \Gamma(t) + i\Delta(t),$$
$$tr_B\{b_i^j b_k^l(t) \rho_B\} = \Gamma(t) - i\Delta(t), \tag{41}$$

where

$$\Gamma(t) + i\Delta(t) = \frac{\lambda^2}{\pi}\int_0^\infty J(\omega)\left\{exp(-i\omega t) + coth\left(\frac{\omega}{2k_B T}\right)cos\,\omega t\right\}. \tag{42}$$

Here $J(\omega)$ is the ohmic damping given by $J(\omega) = \theta(\omega_c - \omega)\eta\omega$, $\Gamma$ is the decoherence rate of the qubit system.

We evaluate the reduced-density-operator in the multiplet basis representation [1]

$$\rho(t) = \sum_{a,b}\rho_{ab}(t) e_{ab}, \quad e_{ab} = |a><b|, \quad a,b = 1,2,3,4, \tag{43}$$

where $e_{ab}$ is the multiplet states. The inner product between the multiplet basis is defined by

$$(e_{ab}, e_{cd}) = tr\left[e_{ab}^\dagger e_{cd}\right] = \delta_{ac}\delta_{bd}. \tag{44}$$

Then, from Eqs. (39)-(44), we obtain the matrix component of the reduced-density-operator as

$$\rho_{ab}(t) = V_{ab|cd}(t)\rho(0),$$
$$V_{ab|cd}(t) = exp\left[-it(E_a - E_b)\right]\left\{\delta_{ac}\delta_{bd} - \left[\delta_{bd}\sum_{a'}M_{aa'a'c} - M_{acdb}\right]k(t) - \left[\delta_{ac}\sum_{a'}M_{aa'a'b} - M_{acdb}\right]k^*(t)\right\}$$
(45)

where

$$M_{abcd} = \sum_{i,j}<a|S_i^j|b><c|S_i^j|d>$$
$$= \frac{1}{4}\sum_{i=1,2}\left\{<a|X_i|b><c|X_i|d> + <a|Y_i|b><c|Y_i|d> + <a|Z_i|b><c|Z_i|d>\right\}, \tag{46}$$

and
$$k(t) = \frac{2}{\pi}\Gamma_o\left\{\frac{\pi}{2}\omega_c t + \int_0^{\frac{t}{\tau_s}} Si(\omega_c \tau_s t)\, dt\right\} \quad (47)$$

$$+ i\Delta_o\left\{\frac{t}{\tau_s} - \frac{1}{\omega_c \tau_s}\left(\frac{\pi}{2}\omega_c \tau_s + Si(\omega_c t)\right)\right\}.$$

Here $\omega_c$ is the high frequency cutoff, $\tau$ is the switching time, $\Gamma_0 = \lambda^2 \eta k_B T \tau_s$ and $\Delta_o = \frac{\lambda^2 \eta \omega_c \tau_s}{\pi}$.

We now study the non-Markovian errors associated with two-qubit gate operations. Here, $Si(x) = \int_0^x dt \frac{\sin t}{t} - \frac{\pi}{2}$, is a sine integral. For reference, we also note the cosine integral, $Ci(x) = -\int_x^\infty dt \frac{\cos t}{t}$. Figure 1 shows the plots of sine and cosine integrals for $0 < \omega_c t < 1$, respectively.

Derivation of Equations (45) to (47) is given in the below:

We define the operator integrand $\widehat{B}$ as follows:

$$\begin{aligned}\widehat{B} &= \widehat{L}_{int}\widehat{U}_o(s,\tau)\widehat{L}_{int}\widehat{U}_o^{-1}(s,\tau)\rho_B\widehat{U}_S^{-1}(s,t)\widehat{U}_S(t)\rho(0)\\
&= \widehat{L}_{int}\widehat{U}_o(s,\tau)\widehat{L}_{int}\widehat{U}_o^{-1}(s,\tau)\rho_B\widehat{U}_S(s)\rho(0)\\
&= \widehat{L}_{int}\widehat{U}_S(s,\tau)\widehat{U}_B(s,\tau)\widehat{L}_{int}\widehat{U}_B(\tau,s)\widehat{U}_S(\tau,s)\rho_B\widehat{U}_S(s)\rho(0)\\
&= \widehat{L}_{int}\widehat{U}_S(s,\tau)\widehat{U}_B(s,\tau)\widehat{L}_{int}\rho_B(s-\tau)\rho(-\tau).\end{aligned} \quad (48)$$

Here, we used the relation

$$\begin{aligned}&\widehat{U}_o^{-1}(s,\tau)\rho_B\widehat{U}_S(s)\rho(0)\\
&= \widehat{U}_o(\tau,s)\rho_B\widehat{U}_S(s)\rho(0)\\
&= \exp[-i\widehat{L}_o(\tau-s)]\rho_B\exp[-is\widehat{L}_S]\rho(0)\\
&= \exp[-i\widehat{H}_o(\tau-s)]\rho_B\exp[-is\widehat{H}_S]\rho(0)\exp[is\widehat{H}_S]\exp[i\widehat{H}_o(\tau-s)]\\
&= \exp[-i\widehat{H}_S(\tau-s)]\exp[-i\widehat{H}_B(\tau-s)]\rho_B\exp[-is\widehat{H}_S]\rho(0)\exp[is\widehat{H}_S]\exp[i\widehat{H}_B(\tau-s)]\exp[i\widehat{H}_S(\tau-s)]\\
&= \exp[-i\widehat{H}_B(\tau-s)]\rho_B\exp[i\widehat{H}_B(\tau-s)]\exp[-i\widehat{H}_S(\tau-s)]\exp[-is\widehat{H}_S]\rho(0)\exp[is\widehat{H}_S]\exp[i\widehat{H}_S(\tau-s)]\\
&= \exp[-i\widehat{H}_B(\tau-s)]\rho_B\exp[i\widehat{H}_B(\tau-s)]\exp[-i\tau\widehat{H}_S]\rho(0)\exp[i\tau\widehat{H}_S]\\
&= \rho_B(s-\tau)\rho(-\tau).\end{aligned}$$

$$(49)$$

In Eq. (49), we have used Baker-Campbell-Hausdorff formula

$$exp(\lambda\widehat{L})\psi = exp(\lambda\widehat{H})\psi exp(-\lambda\widehat{H}),\ \widehat{L}\psi = [\widehat{H},\psi]. \quad (50)$$

Equation (48) is further expanded as

$$\widehat{B} = \widehat{L}_{int} \widehat{U}_o(s,\tau) \widehat{L}_{int} \rho_B(s-\tau) \rho(-\tau)$$
$$= \widehat{L}_{int} e^{-i\widehat{L}_o(s-\tau)} [\widehat{H}_{int}, \rho_B(s-\tau)\rho(-\tau)]$$
$$= \widehat{L}_{int} e^{-i\widehat{H}_o(s-\tau)} [\widehat{H}_{int}, \rho_B(s-\tau)\rho(-\tau)] e^{i\widehat{H}_o(s-\tau)} \quad (51)$$
$$= \widehat{L}_{int} [\widehat{H}_{int}(\tau-s), \rho_B \rho(-s)]$$
$$= [\widehat{H}_{int}, [\widehat{H}_{int}(\tau-s), \rho_B \rho(-s)]].$$

Substituting Eq. (40) and take a trace over the environment, we get

$$tr_B(\widehat{B})$$
$$= \lambda^2 \sum_{i,j,k,l} tr_B(b_i^j b_k^l(\tau-s)\rho_B) \{S_i^j S_k^l(\tau-s)\rho(-s) - S_k^l(\tau-s)\rho(-s) S_i^j\}$$
$$+ \lambda^2 \sum_{i,j,k,l} tr_B(b_k^l(\tau-s) b_i^j \rho_B) \{\rho(-s) S_k^l(\tau-s) S_i^j - S_i^j \rho(-s) S_k^l(\tau-s)\}$$
$$= \lambda^2 \sum_{i,j,k,l} tr_B(b_i^j b_k^l(\tau-s)\rho_B) [S_i^j, S_k^l(\tau-s)\rho(-s)]$$
$$+ \lambda^2 \sum_{i,j,k,l} tr_B(b_k^l(\tau-s) b_i^j \rho_B) [\rho(-s) S_k^l(\tau-s), S_i^j]$$
$$= \sum_{i,j} \{[S_i^j, S_k^l(\tau-s)\rho(-s)](\Gamma(\tau-s) - i\Delta(\tau-s)) + [\rho(-s) S_k^l(\tau-s), S_i^j](\Gamma(\tau-s) + i\Delta(\tau-s))\}.$$
$$(52)$$

Then, we obtain

$$\widehat{V}^{(2)}(t)\rho(0)$$
$$= \widehat{U}_S(t)\rho(0) - \widehat{U}_S(t) \int_0^t ds \int_0^s d\tau \widehat{U}_S^{-1}(s) \sum_{i,j} [S_i^j, S_i^j(\tau-s)\rho(-s)](\Gamma(\tau-s) - i\Delta(\tau-s))$$
$$- \widehat{U}_S(t) \int_0^t ds \int_0^s d\tau \widehat{U}_S^{-1}(s) \sum_{i,j} [\rho(-s) S_i^j(\tau-s), S_i^j](\Gamma(\tau-s) + i\Delta(\tau-s)).$$
$$(53)$$

If we evaluate $\widehat{V}^{(2)}(t)$ in the multiplet basis, we obtain
$$\widehat{V}^{(2)}{}_{ab|cd} = (e_{ab}, \widehat{V}^{(2)}(t) e_{cd}). \quad (54)$$

Let's define operator $\widehat{\Xi}_1, \widehat{\Xi}_2$ as follows:
$$\widehat{\Xi}_1 = \widehat{U}_S^{-1}(s) \sum_{i,j} [S_i^j, S_i^j(\tau-s)\rho(-s)] = \sum_{i,j} [S_i^j(s), S_i^j(\tau)\rho(0)],$$
$$\widehat{\Xi}_2 = \widehat{U}_S^{-1}(s) \sum_{i,j} [\rho(-s) S_i^j(\tau-s), S_i^j] = \sum_{i,j} [\rho(0) S_i^j(\tau), S_i^j(s)].$$

We first calculate $(e_{ab}, \widehat{\Xi}_1)$ which becomes

$$(e_{ab}, \widehat{\Xi}_1)$$
$$= \left(e_{ab}, \sum_{i,j} \sum_{c,d} [S_i^j(s), S_i^j(\tau) e_{cd}] \rho_{cd}(0)\right) \quad (55)$$
$$= \sum_{cd} \left\{\delta_{ab} \sum_{a'} M_{aa'a'c} e^{i\tau(E_{a'} - E_c) + is(E_a - E_{a'})} - M_{acdb} e^{i\tau(E_a - E_c) + is(E_d - E_b)}\right\}.$$

Likewise, $(e_{ab}, \widehat{\Xi}_2)$ becomes

$$\left(e_{ab},\widehat{\Xi}_2\right)$$
$$=\left(e_{ab},\sum_{i,j}\sum_{c,d}\left[e_{cd}S_i^j(\tau),S_i^j(s)\right]\rho_{cd}(0)\right) \tag{55}$$
$$=\sum_{cd}\left\{\delta_{ac}\sum_{a'}M_{da'a'b}e^{i\tau(E_d-E_{a'})+is(E_{a'}-E_b)}-M_{acdb}e^{i\tau(Ed-Eb)+is(E_a-E_c)}\right\}.$$

where

$$M_{abcd}=\sum_{i,j}<a|S_i^j|b><c|S_i^j|d>$$
$$=\frac{1}{4}\sum_{i=1,2}\left\{<a|X_i|b><c|X_i|d>+<a|Y_i|b><c|Y_i|d>+<a|Z_i|b><c|Z_i|d>\right\}. \tag{56}$$

We define $K_{ab|cd}(s,\tau)$ as

$$K_{ab|cd}(s,\tau)$$
$$=\left\{\delta_{ab}\sum_{a'}M_{aa'a'c}e^{i\tau(E_{a'}-E_c)+is(E_a-E_{a'})}-M_{acdb}e^{i\tau(E_a-E_c)+is(E_d-E_b)}\right\}(\Gamma(\tau-s)-i\Lambda(\tau-s))$$
$$+\left\{\delta_{ac}\sum_{a'}M_{da'a'b}e^{i\tau(E_d-E_{a'})+is(E_{a'}-E_b)}-M_{acdb}e^{i\tau(Ed-Eb)+is(E_a-E_c)}\right\}(\Gamma(\tau-s)+i\Lambda(\tau-s)).$$
$$\tag{57}$$

By integrating Eq. (57) over $ds,d\tau$, we obtain

$$\int_0^t ds\int_0^s d\tau K_{ab|cd}(s,\tau)$$
$$=\delta_{ab}\sum_{a'}M_{aa'a'c}k_{a'a'|ca}(t)-M_{acdb}k_{ab|cd}(t)+\delta_{ac}\sum_{a'}M_{da'a'b}k^*_{a'a'|db}(t)M_{acdb}k^*_{ba|dc}(t), \tag{58}$$

$$k_{ab|cd}(t)=\int_0^t dse^{is(\omega_{ac}-\omega_{bd})}\int_0^s d\tau e^{-i\tau\varpi_{ac}}(\Gamma(\tau)+i\Lambda(\tau)),$$

$$\omega_{ab}=E_a-E_b.$$

Equation (42) becomes

$$\Gamma(t)+i\Lambda(t)=\frac{\lambda^2}{\pi}\int_0^\infty J(\omega)\left\{\exp(-i\omega t)+\coth\left(\frac{\omega}{2k_BT}\right)\cos\omega t\right\}$$
$$=\frac{2\lambda^2\eta k_BT}{\pi}\frac{\sin(\omega_c t)}{t}-i\frac{\lambda^2\eta\omega_c}{\pi}\left[\frac{\sin(\omega_c t)}{\omega_c t^2}-\frac{\cos(\omega_c t)}{t}\right] \tag{59}$$
$$=\frac{\pi\Gamma_o}{2\tau_s}\frac{\sin(\omega_c t)}{t}-i\frac{\Delta_o}{\tau_s}\left[\frac{\sin(\omega_c t)}{\omega_c t^2}-\frac{\cos(\omega_c t)}{t}\right].$$

In order to evaluate the integrals in (58) and (59), we need to calculate the integral

$$I_2(s)=\int_0^s d\tau\frac{\sin(\omega'_c\tau)}{\tau}, \text{ which is given by}$$

$$I_2(s) = \frac{\pi \omega'_c}{2} - \sin(\omega'_c s) \int_0^\infty dx \frac{xe^{-xs}}{x^2 + \omega'^2_c} - \omega'_c \cos(\omega'_c s) \int_0^\infty dx \frac{e^{-xs}}{x^2 + \omega'^2_c}$$
$$= \frac{\pi \omega'_c}{2} + Si(\omega'_c s), \tag{60}$$

$$\omega'_c = \omega_c \tau_s.$$

Here $Si(x) = \int_0^x dt \frac{\sin t}{t} - \frac{\pi}{2}$, is a sine integral and $Ci(x) = -\int_x^\infty dt \frac{\cos t}{t}$ is a cosine integral. In Eq. (60), we have used the following relations [6]:

$$\int_0^\infty dx \frac{xe^{-x\mu}}{x^2 + \beta^2} = -Ci(\beta\mu)\cos(\beta\mu) - Si(\beta\mu)\sin(\beta\mu),$$
$$\int_0^\infty dx \frac{e^{-x\mu}}{x^2 + \beta^2} = \frac{1}{\beta}\left[Ci(\beta\mu)\sin(\beta\mu) - Si(\beta\mu)\cos(\beta\mu)\right]. \tag{61}$$

We now define $I_1(t) = \int_0^t ds\, I_2(s)$, which is given by

$$I_1(t) = \frac{\pi}{2}\omega'_c t + \int_0^t Si(\omega'_c s) ds. \tag{62}$$

From Eqs. (58)-(62), we obtain $k_{ab|cd}(t) = \int_0^t ds \int_0^s d\tau K_{ab|cd}(s,\tau)$ given by

$$k_{ab|cd}(t) = \int_0^t ds \int_0^s d\tau K_{ab|cd}(s,\tau)$$
$$= \frac{2}{\pi}\Gamma_o I_1\left(\frac{t}{\tau_s}\right) + i\Delta_o\left\{\frac{t}{\tau_s} - \frac{1}{\omega'_c} I_2\left(\frac{t}{\tau_s}\right)\right\}$$
$$= \frac{2}{\pi}\Gamma_o\left\{\frac{\pi}{2}\omega_c t + \int_0^{\frac{t}{\tau_s}} Si(\omega_c \tau_s t) dt\right\} \tag{63}$$
$$+ i\Delta_o\left\{\frac{t}{\tau_s} - \frac{1}{\omega_c \tau_s}\left[\frac{\pi}{2}\omega_c \tau_s + Si(\omega_c t)\right]\right\} = k(t).$$

This establishes Eqs. (45) to Eq. (47).

### III. Two-qubit gate operations with non-Markovian noise sources

We consider the CNOT gate operation for various input states.

We now consider the non-Markovian error associated with CNOT gate operation. For CNOT gate operation, the multiplet basis is given by

$$|1>^{CNOT} = |00>,$$
$$|2>^{CNOT} = |01>,$$
$$|3>^{CNOT} = \frac{1}{\sqrt{2}}(|10> + |11>), \quad (64)$$
$$|4>^{CNOT} = \frac{1}{\sqrt{2}}(|10> - |11>).$$

In NISQ machines, the input and out states are represented by the computational basis:
$$e^c_{\alpha\beta} = |\alpha><\beta|, \ \alpha,\beta = 1,2,3,4, \ \{|00>,|01>,|10>,|11>\}. \quad (65)$$

Then the reduce-density-operator in the computational basis is given by
$$\rho^c_{\alpha\beta}(t) = \sum_{a',b'} C_{\alpha\beta|a'b'} \rho^m_{a'b'}(t),$$
$$C_{\alpha\beta|ab} = tr\left(e^{c\dagger}_{\alpha\beta}, e^m_{ab}\right). \quad (66)$$

Case 1: When that the initial state is given by $\rho_{cd}(0) = |00><00| = \rho^{CNOT}_{11}(0)$, we obtain

$$\rho^{CNOT}_{11}(t) = V^{CNOT}_{11|11}(t) \rho^{CNOT}_{11}(0)$$
$$= \left\{1 - 2\left[\sum_{a'} M^{CNOT}_{1a'a'1} - M^{CNOT}_{1111}\right] Re\ k(t)\right\} \rho^{CNOT}_{11}(0)$$
$$= (1 - 2\ Re\ k(t)) \rho^{CNOT}_{11}(0),$$

$$\rho^{CNOT}_{22}(t) = V^{CNOT}_{22|11}(t) \rho^{CNOT}_{11}(0)$$
$$= 2 M^{CNOT}_{2112}\ Re\ k(t) \rho^{CNOT}_{11}(0)$$
$$= Re\ k(t) \rho^{CNOT}_{11}(0),$$

$$\rho^{CNOT}_{33}(t) = V^{CNOT}_{33|11}(t) \rho^{CNOT}_{11}(0) \quad (67)$$
$$= 2 M^{CNOT}_{3113}\ Re\ k(t) \rho^{CNOT}_{11}(0)$$
$$= \frac{1}{2} Re\ k(t) \rho^{CNOT}_{11}(0),$$

$$\rho^{CNOT}_{44}(t) = V^{CNOT}_{44|11}(t) \rho^{CNOT}_{11}(0)$$
$$= 2 M^{CNOT}_{4114}\ Re\ k(t) \rho^{CNOT}_{11}(0)$$
$$= \frac{1}{2} Re\ k(t) \rho^{CNOT}_{11}(0).$$

Using Eqs. (64)-(66) and (67), we obtain the computational basis representation of the reduced-density-operator as

$$\rho_{11}^c(t) = (1 - 2\,Re\,k(t))\,\rho_{11}^{CNOT}(0),$$

$$\rho_{22}^c(t) = Re\,k(t)\,\rho_{11}^{CNOT}(0),$$

$$\rho_{33}^c(t) = \frac{1}{2} Re\,k(t)\,\rho_{11}^{CNOT}(0) \tag{68}$$

$$\rho_{44}^c(t) = \frac{1}{2} Re\,k(t)\,\rho_{11}^{CNOT}(0).$$

Case 2: When that the initial state is given by $\rho_{cd}(0) = |01\rangle\langle 01| = \rho_{22}^{CNOT}(0)$, we obtain

$$\rho_{11}^{CNOT}(t) = V^{CNOT}{}_{11|22}(t)\,\rho_{22}^{CNOT}(0)$$
$$= 2M^{CNOT}{}_{1221}\,Re\,k(t)\,\rho_{22}^{CNOT}(0)$$
$$= Re\,k(t)\,\rho_{22}^{CNOT}(0),$$

$$\rho_{22}^{CNOT}(t) = V^{CNOT}{}_{22|22}(t)\,\rho_{22}^{CNOT}(0)$$
$$= \left\{ 1 - 2\left[ \sum_{a'} M^{CNOT}{}_{2a'a'2} - M^{CNOT}{}_{2222} \right] Re\,k(t) \right\} \rho_{22}^{CNOT}(0)$$
$$= (1 - 2\,Re\,k(t))\,\rho_{22}^{CNOT}(0),$$

$$\rho_{33}^{CNOT}(t) = V^{CNOT}{}_{33|22}(t)\,\rho_{22}^{CNOT}(0) \tag{69}$$
$$= 2M^{CNOT}{}_{3223}\,Re\,k(t)\,\rho_{22}^{CNOT}(0)$$
$$= \frac{1}{2} Re\,k(t)\,\rho_{22}^{CNOT}(0),$$

$$\rho_{44}^{CNOT}(t) = V^{CNOT}{}_{44|22}(t)\,\rho_{22}^{CNOT}(0)$$
$$= 2M^{CNOT}{}_{4224}\,Re\,k(t)\,\rho_{22}^{CNOT}(0)$$
$$= \frac{1}{2} Re\,k(t)\,\rho_{22}^{CNOT}(0).$$

Using Eqs. (64)-(67) and (69), we obtain the computational basis representation of the reduced-density-operator as

$$\rho_{11}^c(t) = Re\,k(t)\,\rho_{22}^{CNOT}(0),$$

$$\rho_{22}^c(t) = (1 - 2\,Re\,k(t))\,\rho_{22}^{CNOT}(0),$$

$$\rho_{33}^c(t) = \frac{1}{2} Re\,k(t)\,\rho_{22}^{CNOT}(0) \tag{70}$$

$$\rho_{44}^c(t) = \frac{1}{2} Re\,k(t)\,\rho_{22}^{CNOT}(0).$$

Case 3: When that the initial state is given by
$$\rho_{cd}(0) = \frac{1}{2}(|10\rangle + |11\rangle)(\langle 10| + \langle 11|) = \rho_{33}^{CNOT}(0), \text{ we obtain}$$

$$\rho_{11}^{CNOT}(t) = V^{CNOT}{}_{11|33}(t)\rho_{33}^{CNOT}(0)$$

$$= 2M^{CNOT}{}_{1331} Re\, k(t)\rho_{33}^{CNOT}(0)$$

$$= \frac{1}{2} Re\, k(t)\rho_{33}^{CNOT}(0),$$

$$\rho_{22}^{CNOT}(t) = V^{CNOT}{}_{22|33}(t)\rho_{33}^{CNOT}(0)$$

$$= 2M^{CNOT}{}_{2332} Re\, k(t)\rho_{33}^{CNOT}(0)$$

$$= \frac{1}{2} Re\, k(t)\rho_{33}^{CNOT}(0),$$

$$\rho_{33}^{CNOT}(t) = V^{CNOT}{}_{33|33}(t)\rho_{33}^{CNOT}(0) \tag{71}$$

$$= \left\{1 - 2\left[\sum_{a'} M^{CNOT}{}_{3a'a'3} - M^{CNOT}{}_{3333}\right] Re\, k(t)\right\}\rho_{33}^{CNOT}(0)$$

$$= \left(1 - \frac{3}{2} Re\, k(t)\right)\rho_{33}^{CNOT}(0),$$

$$\rho_{44}^{CNOT}(t) = V^{CNOT}{}_{44|33}(t)\rho_{33}^{CNOT}(0)$$

$$= 2M^{CNOT}{}_{4334} Re\, k(t)\rho_{33}^{CNOT}(0)$$

$$= \frac{1}{2} Re\, k(t)\rho_{33}^{CNOT}(0).$$

Using Eqs. (64)-(67) and (71), we obtain the computational basis representation of the reduced-density-operator as

$$\rho_{11}^{c}(t) = \frac{1}{2} Re\, k(t)\rho_{33}^{CNOT}(0),$$

$$\rho_{22}^{c}(t) = \frac{1}{2} Re\, k(t)\rho_{33}^{CNOT}(0),$$

$$\rho_{33}^{c}(t) = \frac{1}{2}(1 - Re\, k(t))\rho_{33}^{CNOT}(0), \tag{72}$$

$$\rho_{44}^{c}(t) = \frac{1}{2}(1 - Re\, k(t))\rho_{33}^{CNOT}(0).$$

Case 4: When that the initial state is given by

$$\rho_{cd}(0) = \frac{1}{2}(|10> - |11>)(<10| - <11|) = \rho_{44}^{CNOT}(0), \text{ we obtain}$$

$$\rho_{11}^{CNOT}(t) = V^{CNOT}{}_{11|44}(t)\rho_{44}^{CNOT}(0)$$
$$= 2M^{CNOT}{}_{1441} Re\, k(t) \rho_{44}^{CNOT}(0)$$
$$= \frac{1}{2} Re\, k(t) \rho_{44}^{CNOT}(0),$$

$$\rho_{22}^{CNOT}(t) = V^{CNOT}{}_{22|44}(t)\rho_{44}^{CNOT}(0)$$
$$= 2M^{CNOT}{}_{2442} Re\, k(t) \rho_{44}^{CNOT}(0)$$
$$= \frac{1}{2} Re\, k(t) \rho_{44}^{CNOT}(0),$$

$$\rho_{33}^{CNOT}(t) = V^{CNOT}{}_{33|44}(t)\rho_{44}^{CNOT}(0) \qquad (73)$$
$$= 2M^{CNOT}{}_{3443} Re\, k(t) \rho_{44}^{CNOT}(0)$$
$$= \frac{1}{2} Re\, k(t) \rho_{44}^{CNOT}(0),$$

$$\rho_{44}^{CNOT}(t) = V^{CNOT}{}_{44|44}(t)\rho_{44}^{CNOT}(0)$$
$$= \left\{1 - 2\left[\sum_{a'} M^{CNOT}{}_{4a'a'4} - M^{CNOT}{}_{4444}\right] Re\, k(t)\right\} \rho_{44}^{CNOT}(0)$$
$$= \left(1 - \frac{3}{2} Re\, k(t)\right) \rho_{44}^{CNOT}(0).$$

Using Eqs. (66)-(67) and (73), we obtain the computational basis representation of the reduced-density-operator as

$$\rho_{11}^{c}(t) = \frac{1}{2} Re\, k(t) \rho_{44}^{CNOT}(0),$$
$$\rho_{22}^{c}(t) = \frac{1}{2} Re\, k(t) \rho_{44}^{CNOT}(0),$$
$$\rho_{33}^{c}(t) = \frac{1}{2}(1 - Re\, k(t)) \rho_{44}^{CNOT}(0), \qquad (74)$$
$$\rho_{44}^{c}(t) = \frac{1}{2}(1 - Re\, k(t)) \rho_{44}^{CNOT}(0).$$

We have tried both ibm_guadalupe through IBM Quantum and ionQ through Amazon Braket to compare the theory with the experiment for the non-Markovian errors associated with CNOT operation. The occurrence probabilities for 1,000 iterations are shown in the following table 1(a) to table 1(d):

Table 1(a): Quantum state fluctuations under CNOT operation for initial state |00>

| Initial state: |00> | |00> | |01> | |10> | |11> |
|---|---|---|---|---|
| Ibm_guadalupe | 0.982 | 0.008 | 0.001 | 0.009 |
| IonQ | 0.984 | 0.008 | 0.003 | 0.005 |
| Theory: probability for each iteration at time t | $1 - 2Re\ k(t)$ | $Re\ k(t)$<br>1) $Re\ k(\tau_s^{IBM}) = 8 \times 10^{-3}$<br>2) $Re\ k(\tau_s^{IonQ}) = 8 \times 10^{-3}$ | $\frac{1}{2} Re\ k(t)$ | $\frac{1}{2} Re\ k(t)$ |

From the table 1(a), the decoherence function $Re\ k(t)$ at switching time $\tau_s$ can be estimated for each NISQ machine as, $Re\ k(\tau_s^{IBM}) = 8 \times 10^{-3}$, and $Re\ k(\tau_s^{IonQ}) = 8 \times 10^{-3}$, respectively. The NISQ system parameter $\Gamma_0 \omega_c \tau_s$ is $3.5 \times 10^{-3}$ for both ibm-guadrupe an IonQ.

Table 1(b): Quantum state fluctuations under CNOT operation for initial state |01>

| Initial state: |01> | |00> | |01> | |10> | |11> |
|---|---|---|---|---|
| Ibm_guadalupe | 0.04 | 0.944 | 0.01 | 0.006 |
| IonQ | 0.017 | 0.973 | 0.003 | 0.007 |
| Theory: probability for each iteration at time t | $Re\ k(t)$<br>1) $Re\ k(\tau_s^{IBM}) = 4 \times 10^{-2}$<br>2) $Re\ k(\tau_s^{IonQ}) = 7.43 \times 10^{-2}$ | $1 - 2Re\ k(t)$ | $\frac{1}{2} Re\ k(t)$ | $\frac{1}{2} Re\ k(t)$ |

From the table 1(b), the decoherence function $Re\ k(t)$ at switching time $\tau_s$ can be estimated for each NISQ machine as, $Re\ k(\tau_s^{IBM}) = 4 \times 10^{-2}$, and $Re\ k(\tau_s^{IonQ}) = 1.7 \times 10^{-2}$, respectively. The NISQ system parameter $\Gamma_0 \omega_c \tau_s$ is $1.75 \times 10^{-2}$ for ibm-guadrupe and $7.44 \times 10^{-3}$ for IonQ, respectively.

Table 1(c): Quantum state fluctuations under CNOT operation for initial state $\frac{1}{\sqrt{2}}(|10\rangle+|11\rangle)$

| Initial state: $\frac{1}{\sqrt{2}}(|10\rangle+|11\rangle)$ | \|00> | \|01> | \|10> | \|11> |
|---|---|---|---|---|
| Ibm_guadalupe | 0.012 | 0.014 | 0.538 | 0.436 |
| IonQ | 0.006 | 0.006 | 0.504 | 0.484 |
| Theory: probability for each iteration at time t | $\frac{1}{2} Re\, k(t)$<br>1) $Re\, k(\tau_s^{IBM}) = 2.4\times 10^{-2}$<br>2) $Re\, k(\tau_s^{IonQ}) = 1.2\times 10^{-2}$ | $\frac{1}{2} Re\, k(t)$ | $\frac{1}{2} - \frac{1}{2} Re\, k(t)$ | $\frac{1}{2} - \frac{1}{2} Re\, k(t)$ |

From the table 1(c), the decoherence function $Re\, k(t)$ at switching time $\tau_s$ can be estimated for each NISQ machine as, $Re\, k(\tau_s^{IBM}) = 2.4\times 10^{-2}$, and $Re\, k(\tau_s^{IonQ}) = 1.2\times 10^{-2}$, respectively. The NISQ system parameter $\Gamma_0 \omega_c \tau_s$ is $1.05\times 10^{-2}$ for ibm-guadrupe and $5.25\times 10^{-3}$ for IonQ, respectively.

Table 1(d): Quantum state fluctuations under CNOT operation for initial state $\frac{1}{\sqrt{2}}(|10\rangle-|11\rangle)$

| Initial state: $\frac{1}{\sqrt{2}}(|10\rangle-|11\rangle)$ | \|00> | \|01> | \|10> | \|11> |
|---|---|---|---|---|
| Ibm_guadalupe | 0.019 | 0.007 | 0.532 | 0.442 |
| IonQ | 0.004 | 0.004 | 0.495 | 0.497 |
| Theory: probability for each iteration at time t | $\frac{1}{2} Re\, k(t)$ | $\frac{1}{2} Re\, k(t)$<br>1) $Re\, k(\tau_s^{IBM}) = 1.4\times 10^{-2}$<br>2) $Re\, k(\tau_s^{IonQ}) = 8\times 10^{-3}$ | $\frac{1}{2} - \frac{1}{2} Re\, k(t)$ | $\frac{1}{2} - \frac{1}{2} Re\, k(t)$ |

From the table 1(d), the decoherence function $Re\, k(t)$ at switching time $\tau_s$ can be estimated for each NISQ machine as, $Re\, k(\tau_s^{IBM}) = 1.4\times 10^{-2}$, and $Re\, k(\tau_s^{IonQ}) = 8.0\times 10^{-3}$, respectively. The NISQ system parameter $\Gamma_0 \omega_c \tau_s$ is $6.125\times 10^{-3}$ for ibm-guadrupe and $3.5\times 10^{-3}$ for IonQ, respectively.

1. Ahn, D., Oh, J. H., Kimm, K. & Hwang, S, W. *Time-convolutionless reduced-density-operator theory of a noisy quantum channel: two-bit quantum gate for quantum-information processing*. Phys. Rev. A **61**, 052310 (2000).

2. Ahn, D. Lee, J., Kim, M. S. & S. W. Hwang, *Self-consistent non-Markovian theory of a quantum state evolution for quantum information processing*. Phys. Rev. **A** 66, 012302 (2002).

3. Hall, M. J. W. Canoniacal form of master equations and characterization of non-Markovianity. Phys. Rev. A 89, 042120 (2014).

4. De Vega, I. & Alonso, D. Dynamics of non-Markovian quantum systems. Rev. Mod. Phys. 89, 015001 (2017).

5. Loss, D. & DiVincenzo, D. P. Quantum computation with quantum dots, Phys. Rev. A 57, 120 (1998).

6. Gradshteyn, I. S. & Ryzhik, I. M. *Table of Integrals, Series and Products*. (Academic Press, San Diego, 2007)